\pdfoutput=1

\documentclass[11pt,twoside,a4paper,cmspaper,final,collab]{cms-tdr}
\def\svnVersion{82432:82433}\def\cmsCernNo{2011-171}\def\cmsCernDate{\today}\def\cmsMessage{Submitted to the Journal of High Energy Physics}

\begin{document}\cmsNoteHeader{QCD-10-035}

\hyphenation{had-ron-i-za-tion}
\hyphenation{cal-or-i-me-ter}
\hyphenation{de-vices}
\RCS$Revision: 82418 $
\RCS$HeadURL: svn+ssh://alverson@svn.cern.ch/reps/tdr2/papers/QCD-10-035/trunk/QCD-10-035.tex $
\RCS$Id: QCD-10-035.tex 82418 2011-10-28 18:52:59Z alverson $

     \newcommand{\ch}[2]{#2}

\cmsNoteHeader{QCD-10-035} 
\title{Measurement of the Production Cross Section for Pairs of Isolated Photons in $\Pp\Pp$ collisions at $\sqrt{s}=7\TeV$}

\date{\today}

\abstract{
   The integrated and differential cross sections for the production of pairs
   of isolated photons is measured in proton-proton collisions at a
   centre-of-mass energy of $7\TeV$ with the CMS detector at the LHC. A data
   sample corresponding to an integrated luminosity of $36~\text{pb}^{-1}$ is
   analysed. A next-to-leading-order perturbative QCD calculation is compared
   to the measurements. A discrepancy is observed for regions of the phase
   space where the two photons have an azimuthal angle difference
   $\Delta\varphi\lesssim 2.8$.
   }

\hypersetup{%
pdfauthor={CMS Collaboration},%
pdftitle={Measurement of the Production Cross Section for Pairs of Isolated Photons in pp collisions at sqrt(s) = 7 TeV},%
pdfsubject={CMS},%
pdfkeywords={CMS, physics}}

\maketitle 

\section{Introduction}
\label{sec:intro}

The production of energetic photon pairs in hadronic collisions is a
valuable testing ground of perturbative quantum chromodynamics (pQCD).
The emission of a pair of photons from hard parton-parton scattering
constitutes a particularly clean test of perturbation theory in the
collinear factorisation~\cite{Binoth:1999qq,Bern:2002jx} and
$\kt$ factorisation ~\cite{Saleev:2009et} approaches, as well as
soft-gluon logarithmic resummation techniques~\cite{Balazs:2007hr}.
A comprehensive understanding of photon pair production is also
important as it represents a major background in certain searches for
rare or exotic processes, such as the production of a light Higgs
boson, extra-dimension gravitons, and some supersymmetric states.

This paper presents a measurement of the production cross section for
isolated photon pairs in proton-proton collisions at a centre-of-mass
energy of $7~\text{TeV}$, using the Compact Muon Solenoid
(CMS) detector at the Large Hadron Collider (LHC).  Isolated photons
produced in the hard scattering of quarks and gluons are henceforth
 referred to as \textit{signal photons} and the remaining
photons as \textit{background photons}. A pair of signal photons will
be referred to as a diphoton. The data sample was collected in 2010
and corresponds to an integrated luminosity of $36.0\pbinv$.
Recent diphoton cross-section measurements have been performed by the
D0~\cite{Abazov:2010ah} and
CDF~\cite{PhysRevLett.107.102003,PhysRevD.84.052006} Collaborations
in proton-antiproton collisions at $\sqrt{s} =1.96~\text{TeV}$,
and by the ATLAS Collaboration at the LHC~\cite{atlas2011}.

The CMS detector consists of a silicon pixel and strip tracker
surrounded by a crystal electromagnetic calorimeter (ECAL) and a
brass/scintillator sampling hadron calorimeter (HCAL), all in an axial
$3.8\,\text{T}$ magnetic field provided by a superconducting solenoid
of $6\,\text{m}$ internal diameter. The muon system is composed of
gas-ionization detectors embedded in the steel return yoke of the
magnet.  In addition to the barrel and endcap detectors, CMS has an
extensive forward calorimetry system. A more detailed description of
CMS can be found elsewhere~\cite{:2008zzk}.

In the CMS coordinate system, $\theta$ and $\varphi$ respectively
designate the polar angle with respect to the counterclockwise beam
direction, and the azimuthal angle, expressed in radians throughout
this paper. The pseudorapidity is defined as $\eta =
-\ln \left[\tan\frac{\theta}{2}\right]$.

Distance in the $(\eta, \varphi)$ plane is defined as $R =
\sqrt{(\Delta\eta)^2 + (\Delta\varphi)^2}$. The transverse energy
$\ET$ of a particle is defined as $\ET = E\sin\theta$, where $E$ is
the energy of the particle, and the transverse momentum is $p_T = p
\sin\theta$. The rapidity is defined as
$y=\frac{1}{2}\ln\big[\frac{E+p_z}{E-p_z}\big]$, with $p_z$ being the
longitudinal momentum with respect to the beam axis.

The electromagnetic calorimeter, which plays a major role in this
measurement, consists of nearly 76\,000 lead tungstate crystals. It is
divided into a central part (barrel) covering the region $|\eta| <
1.48$ and forward parts (endcaps) extending the coverage up to
$|\eta| < 3$ for a particle originating from the nominal interaction
point. The crystals are arranged in a projective geometry with a
granularity of $0.0174$ in both the $\eta$ and $\varphi$ directions in
the barrel, and increasing with $\eta$ from $0.021$ to $0.050$ in the
endcaps.  A preshower detector, consisting of two planes of silicon
sensors interleaved with 3 radiation lengths of lead, is
placed in front of the endcaps to cover the pseudorapidity region
$1.65 < |\eta| < 2.6$.

The differential cross section is measured as a function of variables
that are particularly relevant in searches for rare processes or to
characterise QCD interactions (e.g.~\cite{Bern:2002jx}):
\begin{itemize}
\item the diphoton invariant mass, $m_{\Pgg\Pgg}$;
\item the azimuthal angle between the two photons,
  $\Delta\varphi_{\Pgg\Pgg}$;
\item the photon pair transverse momentum,
  $p_{T,\Pgg\Pgg}=\sqrt{{p_{T,\Pgg_1}}^{2}+{p_{T,\Pgg_{2}}}^{2}+2\,p_{T,\Pgg_1}p_{T_,\Pgg_2}\cos\Delta\varphi_{\Pgg\Pgg}}$,
  where $p_{T,\Pgg_1}$ and $p_{T,\Pgg_2}$ are the magnitudes of
  the transverse momenta of the two photons;
\item $|{\cos\theta^*}|=|{\tanh\frac{\Delta y_{\Pgg\Pgg}}{2}}|$,
 with $\Delta y_{\Pgg\Pgg}$ being the difference between the two
  photon rapidities. At lowest order in QCD, $\theta^*$ is the center-of-mass
  scattering angle for the $q\bar{q}\rightarrow\Pgg\Pgg$ and
  $gg\rightarrow\Pgg\Pgg$ processes.
\end{itemize}

The event selection requires at least one isolated photon with $\ET >
23\GeV$ and a second isolated photon with $\ET > 20\GeV$, separated by
$R > 0.45$. The measurements are performed in two pseudorapidity
regions, one with $|\eta| < 1.44$, and the other defined by the
tracker acceptance $|\eta| < 2.5$, but excluding the transition region
between the barrel and endcap calorimeters, $1.44 < |\eta| < 1.57$.
For convenience the widest $\eta$ range without the transition will be
referred to as $|\eta| < 2.5$ throughout the paper.

The asymmetric thresholds on the photon transverse momenta
avoid the infrared sensitivity affecting the fixed-order
calculations~\cite{Frixione:1997ks,Fontannaz:2001nq} and simplify the
comparison of the measurements with the theoretical predictions.

All simulation results are based on the
\PYTHIA~6.4.22~\cite{Sjostrand:2006za} event generator, with the \textsc{z2}
tune, the \textsc{cteq6l} parton distribution functions
(PDFs)~\cite{Pumplin:2002vw}, and a \GEANTfour~\cite{Pia:2003cj}
modelling of the detector. The {\sc z2} tune is identical to the \textsc{z1} tune
described in~\cite{Field:2010bc} except that {\sc z2} uses the
{\sc cteq6l} PDFs while {\sc z1} uses
{\sc cteq5l}~\cite{springerlink:10.1007/s100529900196}. At the generator
level, a prompt photon is considered as signal if the sum of the
generated transverse momenta of all the particles within a cone
$R<0.4$ around the photon direction is less than $5\GeV$.

Event selection and background discrimination are presented in
Sections~\ref{sec:recosel} and \ref{sec:extraction}. The determination
of the signal yield and the measurement of the cross section are
described in Sections~\ref{sec:yield} and
\ref{sec:xsection}. Systematic uncertainties are detailed in Section~\ref{sec:syst}.
 Results are discussed in Section~\ref{sec:results} and compared with the theoretical
predictions introduced in Section~\ref{sec:theory}.

\section{Event Selection}
\label{sec:recosel}

Photon candidates are reconstructed by clustering the energy deposited
in the ECAL~\cite{egm-10-005, egm-10-006} crystals. CMS is equipped
with a versatile trigger to adapt to the steady increase in the LHC
instantaneous luminosity. In this measurement, three trigger settings
were used for three successive data-taking periods. They require two
photon candidates, with a threshold of either $15\GeV$ or $17\GeV$ on
the transverse energy. For the last period, with the highest
instantaneous luminosity, a weak isolation requirement is applied on
one of the two photon candidates. For the three periods, the trigger
efficiency for events passing the analysis selections described in the
following paragraphs is estimated from simulated events to be greater
than 99.9\%.  The offline event selection requires one photon
candidate with $\ET > 23\GeV$ and a second photon candidate with $\ET
> 20\GeV$, each within the fiducial region defined in the
introduction. The candidates are required to be separated by $R >0.45$
to avoid energy deposits related to one candidate overlapping with the
isolation region of another candidate.

Photon identification criteria requiring the deposits in the
calorimeters to be consistent with an electromagnetic shower are
applied to the two candidates. The criteria are based on the spread
along $\eta$ of the energy clustered in the ECAL, henceforth referred
to as $\sigma_{\eta\eta}$, and on the ratio $H/E$ of the energies
measured in the HCAL and ECAL ({\em loose selections} of
Ref.~\cite{egm-10-006}).

The photon candidates are required to be isolated. The sum of the
transverse momenta of charged particles measured by the tracker and
the sum of the transverse energy deposits in the HCAL, both defined
within a cone of radius $R = 0.4$ around the photon direction, must
each be less than $2\GeV$ in the barrel and $4\GeV$ in the endcaps.
HCAL deposits in a cone of radius $R = 0.15$ are excluded from the
sum, as well as tracks in a cone of radius $R = 0.04$ and within a
strip of $\Delta\eta = 0.03$ along the $\varphi$ direction, which can
potentially contain tracks of an electron-positron pair from the
conversion of a photon in the tracker material.  The sum of the
transverse energy deposited in the ECAL in a cone of radius $R=0.3$ is
required to be less than $20\%$ of the photon transverse energy, in
order to be consistent with the online trigger requirements.  Excluded
from the sum is the energy deposited within a cone whose radius
corresponds to 3.5~crystals along $\eta$ and within a 5-crystal-wide
strip extending along the $\varphi$ direction.  In addition, we
require that no charged particle with the following properties impinge
on the ECAL within a cone of radius $R=0.4$: transverse momentum $\pt
> 3\GeV$, impact parameters with respect to the primary vertex in the
transverse and longitudinal planes of less than $1\mm$ and $2\mm$,
respectively, and one associated hit in the innermost layer of the
pixel detector.  Tracks corresponding to such particles are henceforth
called \emph{impinging tracks}. The electron contamination is further
reduced by imposing an additional veto on the presence of hits in the
layers of the pixel detector along the direction of the photon
candidate.

\section{Signal and Background Discrimination}
\label{sec:extraction}

The photon candidates in the selected event sample are designated as
  signal photons, background photons from hadron decays (most of which are
misidentified pairs of collinear photons coming from neutral meson decays),
 or misidentified electrons.
The background to diphoton pair events is thus made up of photon+jet
and multijet events, with respectively one and two background photons
from neutral hadron decays, and Drell--Yan events, with two misidentified electrons.

The contamination from Drell--Yan events is estimated from
simulation using the next-to-leading-order (NLO) \textsc{powheg}
generator~\cite{Alioli:2010xd,Frixione:2007vw,Nason:2004rx}, which
agrees well with our own Drell--Yan measurement~\cite{Chatrchyan:2011cm}.  The
diphoton cross-section measurement is corrected for this
contamination, which amounts to about $12\%$ in the diphoton mass range
$80$--$100\GeV$ around the $\Z$ peak. This procedure has a negligible
impact on the systematic uncertainties.

Background photons from photon+jet and multijet events are produced in
jets alongside other particles, which tend to widen the deposits in
the ECAL.
An isolation variable $\mathcal{I}$ based on the energy in the
ECAL is used to statistically estimate the fraction of diphoton events
among the selected candidates. This variable is constructed to
minimise the dependence on the energy deposited by minimum-ionising
particles (MIPs) such that its distribution for the background can be
obtained from the data by means of the impinging-track method
described below.
It is defined as the sum of the transverse energy of the ECAL deposits
with $\ET>300~\text{MeV}$ (MIP veto), within a hollow cone centred on
the photon impact point, with an inner radius of $3.5$ crystal widths and
an outer radius of $R = 0.4$. Deposits assigned to the photon itself
or falling within a $5$-crystal-wide strip extending along $\varphi$
and centred on the photon impact point are removed. Thus, deposits
from photons converting into electron-positron pairs in the tracker
material and spread along the $\varphi$ direction do not contribute to the value of
$\mathcal{I}$.
The variable $\mathcal{I}$ differs from the ECAL isolation used
in the selection described in Section~\ref{sec:recosel}.

\begin{figure}[p]
  \centering
    \includegraphics[width=0.47\textwidth] {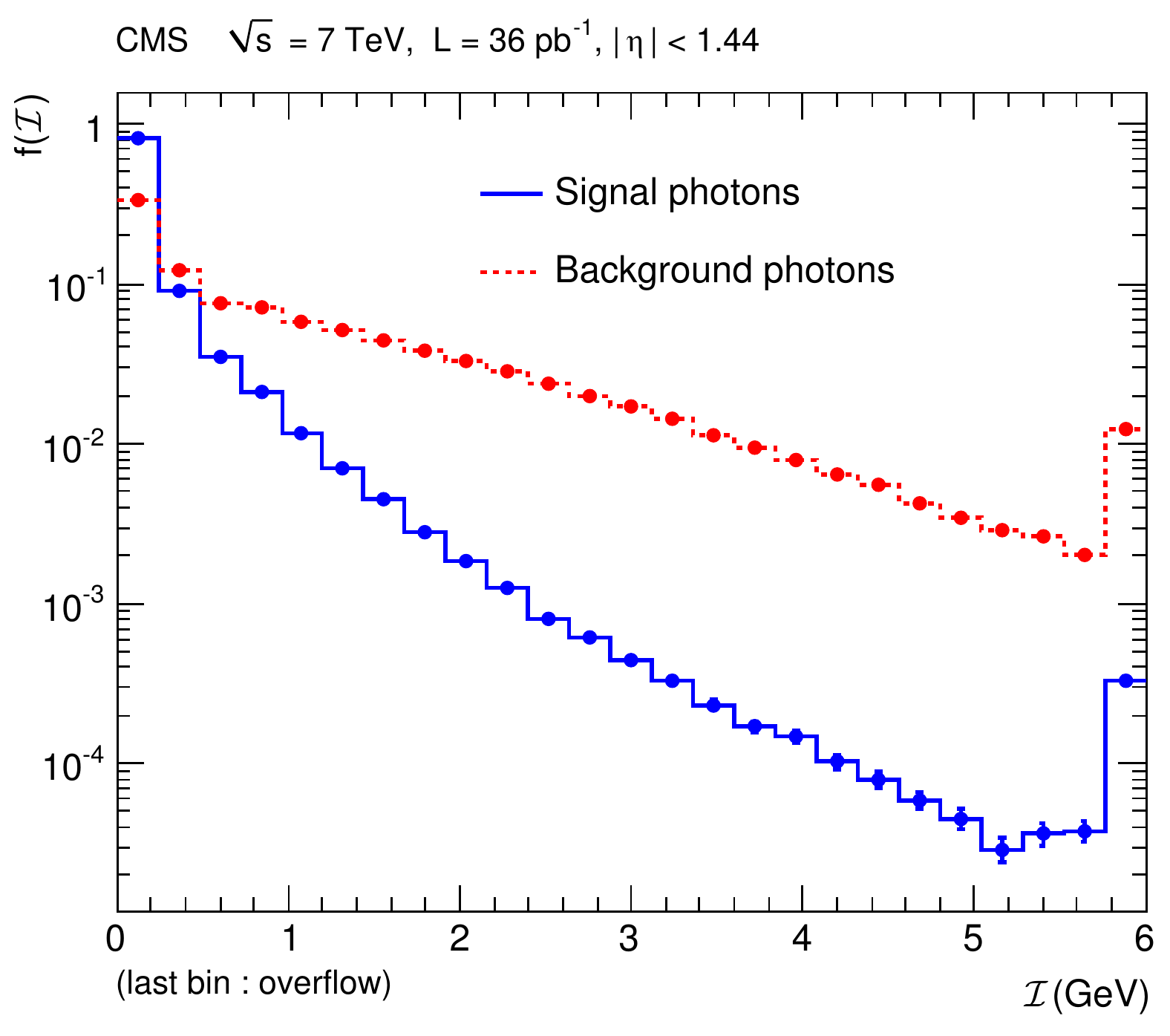}
    \includegraphics[width=0.47\textwidth] {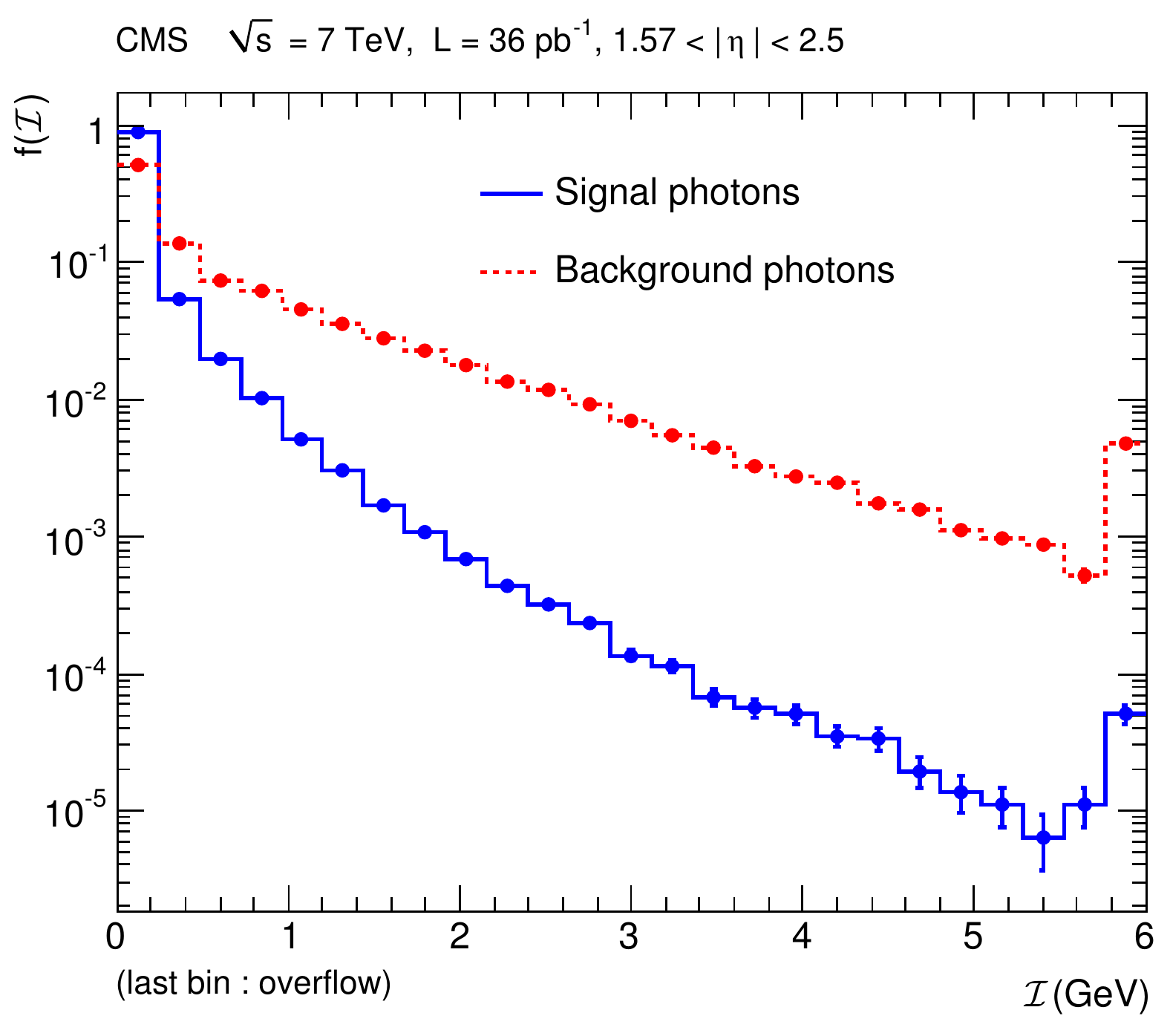}
    \caption{Probability density functions of the ECAL isolation
      variable $\mathcal{I}$ for signal photons (solid blue) and
      background photons (dashed red) in the barrel (left) and in the
      endcap regions (right).  }
  \label{fig:pdfs}
\end{figure}

As the distribution of $\mathcal{I}$ is different for signal
photons and background photons, this variable can be used in a
maximum-likelihood fit to extract the number of signal events in the entire
selected sample. Figure~\ref{fig:pdfs} shows the probability density
function of $\mathcal{I}$, which was extracted from data with the
methods described below.

Contributions to the value of $\mathcal{I}$ for signal photons come
from pileup (multiple proton-proton collisions in the same bunch
crossing) and underlying-event activity (multiple parton interaction
and beam remnants from the same proton-proton collision). Since these
contributions are independent of $\varphi$, the ECAL isolation
probability density function $f(\mathcal{I})$ is estimated from
\textit{random cones} using events with at least one isolated photon
candidate. The value of $\mathcal{I}$ is calculated in a cone of
radius $R = 0.4$ around an axis at the same value of $\eta$ as the photon
candidate and at a random value of $\varphi$ within a window of width
$\pi/2$ centred on the axis perpendicular to the photon direction,
with the same exclusions applied to photon signals. The cone is
required not to include any photon or electron candidates or jets.
The function $f(\mathcal{I})$ for signal photons is
validated with two additional independent methods. Both methods
exploit $\Pep$ and $\Pem$ from $\Z$ and $\PW$ boson decays that do
not radiate significantly in the tracker material. The $\Pep$ and
$\Pem$ are selected with a constraint imposed on the fraction of
bremsstrahlung energy emitted from the interaction in the tracker
material.
Such electrons and positrons leave
ECAL energy deposits consistent with those of photons, and have a
similar probability density function for $\mathcal{I}$. The
$\Z \rightarrow \Pep\Pem$ events are selected with stringent requirements
on the identification criteria of the lepton pair and on its invariant
mass, and the $f(\mathcal{I})$ distribution is obtained directly from
both leptons. In $\PW \rightarrow \Pe\Pgne$ events, $f(\mathcal{I})$ is
obtained by exploiting the \emph{sPlot} technique~\cite{ledib}. The
missing transverse energy projected along the lepton axis
is used to estimate the probability of an event to be
signal ($\PW \rightarrow \Pe\Pgne$) or background ($\Z \rightarrow \Pep\Pem$,
$\PW \rightarrow\Pgt\Pgngt$, $\Pgg + \text{jet(s)}$, and QCD multijet
processes). The value of $\mathcal{I}$ for the selected candidates
is weighted accordingly to estimate the distribution of
$\mathcal{I}$.  The uncertainty on $f(\mathcal{I})$ is taken as the
maximum difference between the distributions extracted from random
cones and from electrons in $\Z$ and $\PW$ events. In simulated events,
the difference between $f(\mathcal{I})$ for signal photons and for
random cones is smaller than the uncertainty determined from data.

For background photons, $f(\mathcal{I})$ is extracted from a
 sample with less than $0.1\%$ of signal-photon
contamination.  The sample is obtained by selecting photon candidates
with one and only one impinging track. A cone of radius $R = 0.05$
around the track is excluded from the isolation area to avoid counting
the energy deposited by the charged particle.  The isolation variable
$\mathcal{I}$ is then rescaled to take into account this additional
exclusion.
To validate this method, the $\mathcal{I}$  distribution is also
extracted from a sample of events with two impinging tracks, one of
the two being excluded in the computation of $\mathcal{I}$. The latter
distribution is compared to that obtained with the
one-impinging-track sample, using the normal definition of
$\mathcal{I}$, i.e., including the energy deposits in the vicinity of
the track.  The agreement is within one standard deviation for the
entire range of the $\mathcal{I}$ distribution, and the difference is
taken as a systematic uncertainty on  $f(\mathcal{I})$ for background
photons.

The distributions $f(\mathcal{I})$ show a moderate dependence on
$\eta$ and on the pileup conditions, the latter being quantified by
the number $n_{\text{vtx}}$ of primary vertices in the events (2.4 on
average).  The background distribution $f(\mathcal{I})$ also depends
on the transverse energy $\ET$ of the candidate. Therefore, events in
the sample used for the extraction of $f(\mathcal{I})$ are weighted to
reproduce the distributions of $\eta$, $n_{\text{vtx}}$, and $\ET$ of
the diphoton sample used for the cross section measurement. The effect
of using the distributions from the diphoton sample to correct the
biases in the background and signal shapes used in the
maximum-likelihood fit is addressed in the systematic uncertainty
section, Section~\ref{sec:syst}.

\section{Signal Yield Determination}
\label{sec:yield}

The number of diphoton events is obtained from a binned maximum-likelihood
fit to the distributions of the ECAL isolation variables of the two
photons, $\mathcal{I}_{1}$ and $\mathcal{I}_{2}$, whose ordering is
chosen randomly. Events are separated into three types: signal events
($\Pgg \Pgg$) if both photons are signal photons, background events
with a signal photon and a background photon, and background events
with two background photons.

The likelihood function $\mathcal{L}$ that is maximised in the fit is

\begin{equation}
  \mathcal{L} =
  \frac{e^{-N^{\text{tot}}}}{N!} \prod_{i=1}^{N}\ \sum_{t=1}^{3} {N_{t}f_{t}(\mathcal{I}^{i}_1,\mathcal{I}^{i}_2})~,
\end{equation}

where  $N$ is the number of selected events, $N_t$ is the number of
events estimated in the fit for event type $t$,
$N^{\text{tot}}$ is their sum, and
$f_t(\mathcal{I}_{1},\mathcal{I}_{2})$ is the probability for the ECAL
isolation variables of the two photons to have values
$\mathcal{I}_{1}$ and $\mathcal{I}_{2}$ for a given event type $t$.

The probability density functions $f_{t}(\mathcal{I}^{i}_1,\mathcal{I}^{i}_2)$
 for the three event types are
obtained by multiplying the probability density functions
$f(\mathcal{I})$ for single-photon candidates, assuming the two
statistical variables $\mathcal{I}_{1}$ and $\mathcal{I}_{2}$ to be
independent.  Correlations between these two variables have been
checked with simulation and are negligible.

\begin{figure}[p]
  \centering
    \includegraphics[width=0.47\textwidth] {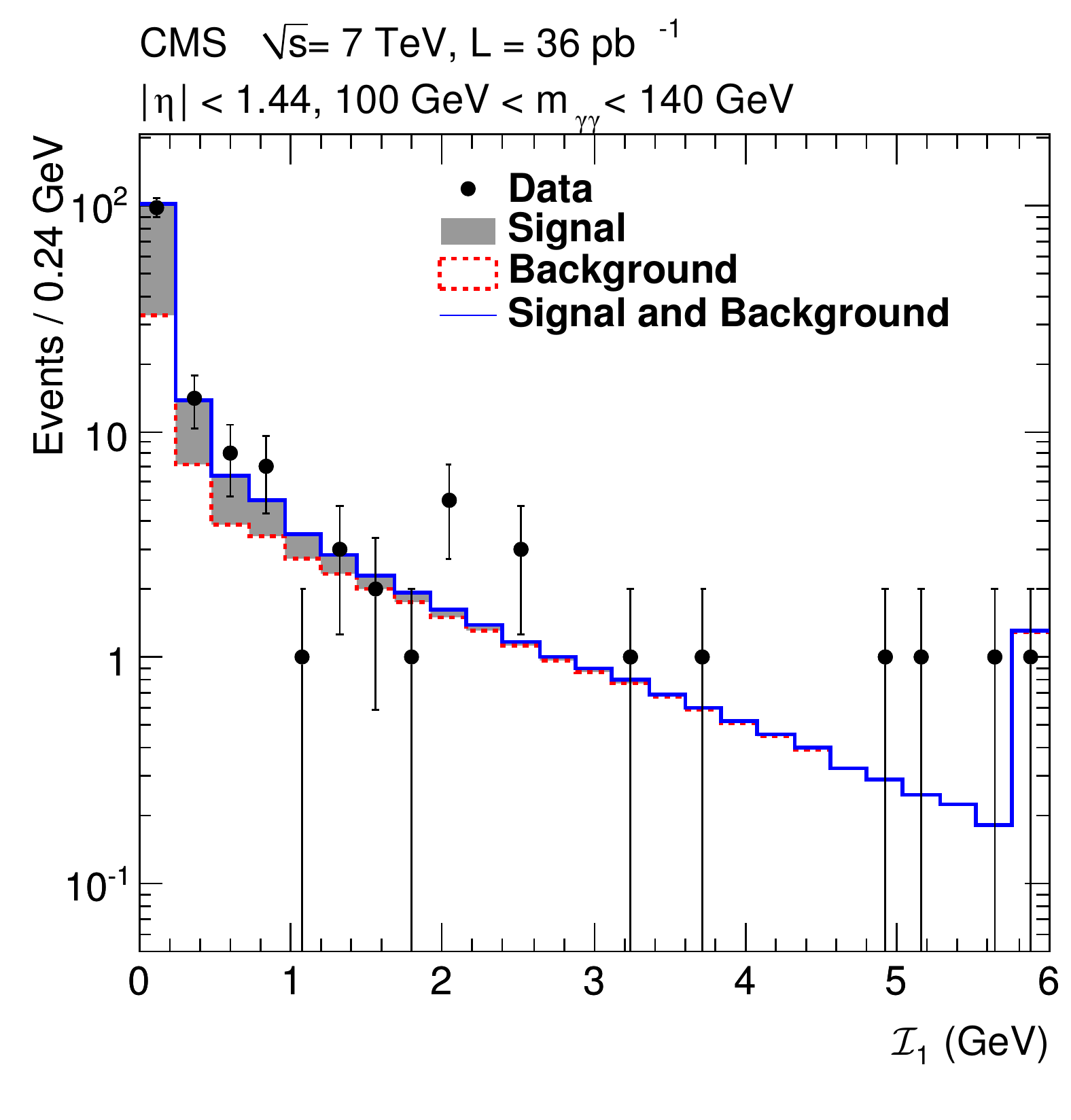}
    \includegraphics[width=0.47\textwidth] {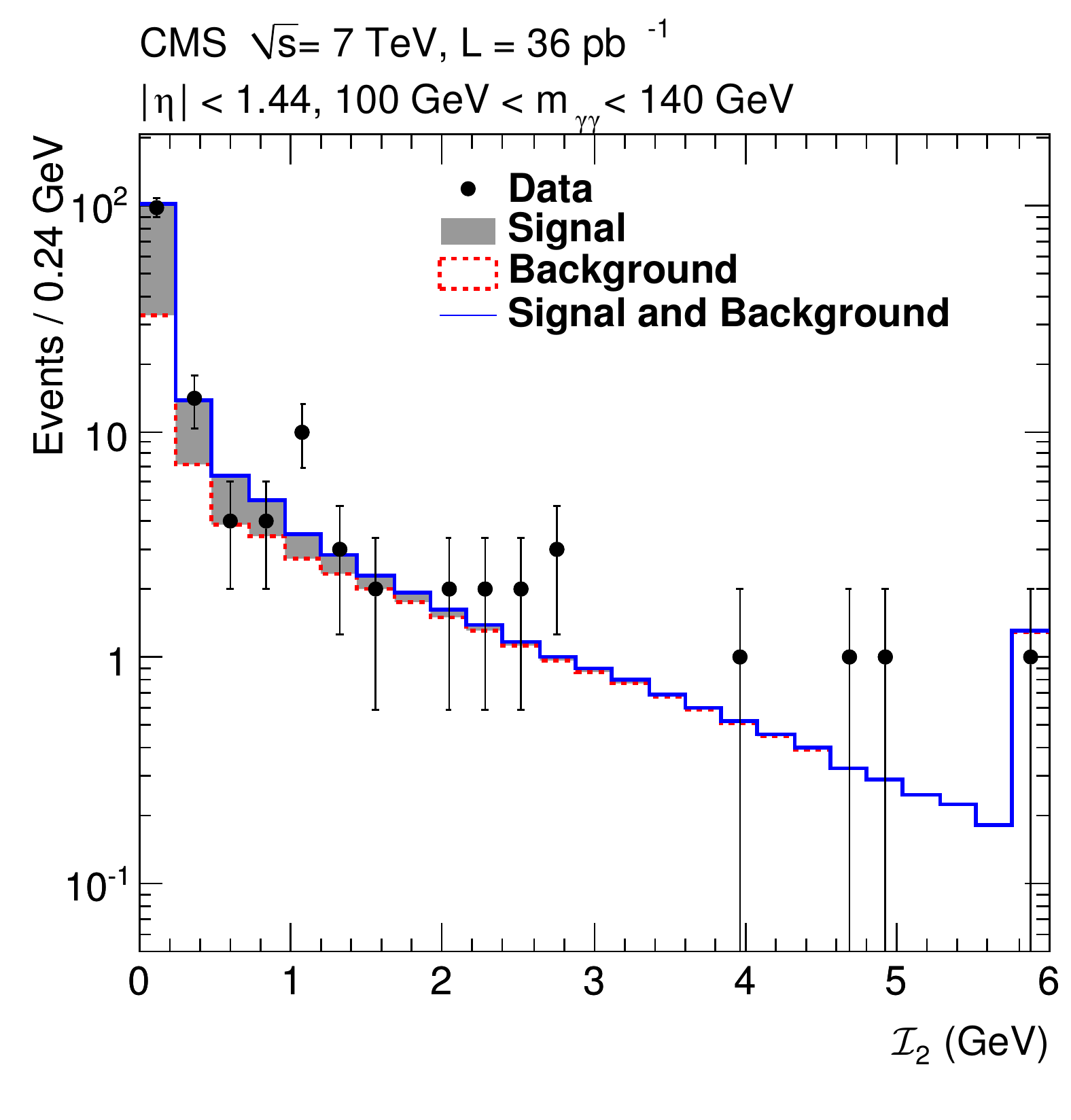}
    \caption{\label{fig1:2dfit} Fit to the photon ECAL isolation
      $(\mathcal{I}_1,\mathcal{I}_2)$ in the bin $100 < m_{\Pgg\Pgg} <
      140\GeV$ for photons with $|\eta| < 1.44$.  The distribution of
      the isolation variable $\mathcal{I}_1$ of one photon candidate,
      arbitrarily chosen as the ``first photon'' and denoted with
      subscript ``1'', is displayed in the left figure, together with
      the fit result, integrated over $\mathcal{I}_2$; the shaded
      region shows the signal distribution, the dashed line represents
      the background contribution, while the solid line is the sum of
      the signal and background contributions. The same distributions
      for the second photon candidate are shown in the right
      figure. In this mass bin, the number of signal events is $72 \pm
      14$, out of the total number of $161$ selected candidates.}
  \label{fig:2dfit}
\end{figure}

A total of 5977 events pass the selection criteria described in
Section~\ref{sec:recosel}.  These events are divided into three
subsamples depending on whether both photons are in the barrel (2191
events), one is in the barrel and the other in the endcaps (2527
events), or both are in the endcaps (1259 events). The fit is
performed separately for each of the three subsamples and each bin of
the four observables. An example of the fit for one bin of the
$m_{\Pgg\Pgg}$ spectrum is shown in Fig. \ref{fig:2dfit} for events
with both photons in the barrel ($|\eta| < 1.44$).

The maximum-likelihood method is known to be biased for samples with
small numbers of events~\cite{Eadie:1977}. This bias is estimated with Monte Carlo
pseudo-experiments and the fit results are corrected for it. It is
less than 10\% of the statistical uncertainty for 80\% of the fits and
never exceeds half the statistical uncertainty.

\section{Cross-Section Measurement}
\label{sec:xsection}

The differential diphoton cross-section measurement $d\sigma/dX$, for the
variable $X$ in the interval $X_i$, is

\begin{equation}
\frac{d\sigma}{dX}(X_i) = \frac{ N_{\Pgg\Pgg}^{\text{U}} (X_i) }{\mathcal{L} \Delta X_i \mathcal{C}(X_i)  }~,
  \label{eq:xsec}
\end{equation}

where $N_{\Pgg\Pgg}^{\text{U}}$ is the number of signal events
obtained from the fit, unfolded for the detector
resolution and corrected for the Drell--Yan contamination;
$\mathcal{L}$ is the integrated luminosity, $\Delta X_i$ is the
interval width; $\mathcal{C}$ is a correction factor for the
effects of the detector resolution on the acceptance and on the
efficiencies of photon reconstruction and identification.

The number of signal events is unfolded~\cite{Cowan:1998ch11} for the
detector resolution by inverting a response matrix $T$  for each of
 the observables $m_{\Pgg\Pgg}$, $p_{T,\Pgg\Pgg}$, $\Delta\varphi_{\Pgg\Pgg}$, and
$|{\cos\theta^*}|$, obtained from simulated events passing the selection requirements.
The matrix elements $T^{ik}$ are the probabilities
of a selected event with the generated value of $X$ in bin $X_k$ to be
reconstructed with a value of $X$ in bin $X_i$. For a given interval
$X_k$, the number of events after unfolding is related to the observed
numbers of events in the different intervals $X_i$ by
$N_{\Pgg\Pgg}^{\text{U}}(X_k) = \big( T^{-1} \big)^{ki}
N_{\Pgg\Pgg}(X_i)$.  Here, $N_{\Pgg\Pgg}(X_i)$ is the signal yield
corrected for the Drell--Yan contamination, as described in
Section~\ref{sec:extraction}.  Given the excellent energy resolution
of the ECAL and the bin sizes, the matrix $T$ is nearly diagonal, and
thus no regularisation is applied in the unfolding procedure.

The correction factor $\mathcal{C}(X_i)$ is defined as
\begin{equation}
  \mathcal{C}(X_i) =
  \frac{N^{\text{sim}}_{\text{reco}}(X_i)}{N^{\text{sim}}_{\text{gen}}(X_i)} %
  \frac{\varepsilon^{\text{data}}}{\varepsilon^{\text{sim}}}, %
  \label{eq:acc}
\end{equation}

where

\begin{description}
\item[$N^{\text{sim}}_{\text{reco}}(X_i)$] is the number of simulated
  events passing all the selection criteria, with generated values of
  $X$ within the interval $X_i$;
\item[$N^{\text{sim}}_{\text{gen}}(X_i)$] is the number of simulated
  events within the acceptance defined at the generator level
  (Section~\ref{sec:intro}), with generated values of $X$ within the
  interval $X_i$;
\item[$\varepsilon^{\text{data}}$] is the efficiency of the photon
  identification criteria measured from data;
\item[$\varepsilon^{\text{sim}}$] is the efficiency of the photon
  identification criteria obtained from simulated events using the same
  technique as for $\varepsilon^{\text{data}}$.
\end{description}

The efficiencies $\varepsilon^{\text{data}}$ and
$\varepsilon^{\text{sim}}$ to observe a diphoton
candidate are taken as the square of the efficiencies to observe a
single photon.

The efficiency for the requirements on isolation, $\sigma_{\eta\eta}$,
and $H/E$ is estimated with a ``tag-and-probe''
method~\cite{Khachatryan:2010xn} applied to a $\Z\rightarrow \Pep\Pem$
sample selected from the full 2010 dataset.  One
lepton, the tag, is selected with tight reconstruction and
identification criteria~\cite{egm-10-004}, while the other, the
probe, is selected by requiring a constraint on the invariant mass of
the lepton pair.  The probes constitute a sample of unbiased electrons
and positrons.  The same constraint as discussed in
Section~\ref{sec:extraction} is applied on the fraction of
bremsstrahlung energy emitted by the $\Pep$ and $\Pem$ interacting in
the tracker material.  This requirement ensures that the
electromagnetic deposits of these electrons and
positrons are consistent with those of a photon shower. The efficiency
is computed by applying the requirements on isolation,
$\sigma_{\eta\eta}$, and $H/E$ to this sample, and then measuring the
fraction of probes passing the selection.

The efficiency for the requirement to have no impinging tracks within
the isolation cone is estimated from data, from a control sample
built using a \textit{random-cone} technique on events with a single
photon selected according to the identification criteria described
above. The random-cone definition is that introduced in
Section~\ref{sec:extraction} for the extraction of $f(\mathcal{I})$.
Particles within the random cone hence come mainly from pileup and the
underlying event. Quantities such as the number of impinging tracks or
energy deposits in the isolation area are therefore assumed
to have the same distributions as for isolated photons. The efficiency of the
requirement to have no impinging track within the isolation cone is
given by the ratio of the number of random cones passing this
criterion to the total number of random cones.  The efficiency of the
veto on pixel hits is obtained from simulation. It is included in the
$N^{\text{sim}}_{\text{reco}}/N^{\text{sim}}_{\text{gen}}$ term of
Eqn.~(\ref{eq:acc}).

The correction factor $\mathcal{C}$ is $(80.8\pm 1.9) \%$ for the
integrated cross section in the region $|\eta|<1.44$, and $(76.2
\pm 3.3) \%$ in the region $|\eta| < 2.5$.

\section{Systematic Uncertainties}
\label{sec:syst}

The systematic uncertainty on the reconstructed photon four-momenta
 is dominated by the ECAL energy scale, known to
$0.6\%$ in the barrel and  $1.5\%$ in the
endcaps~\cite{cms-dps-2011-008}. The energy scale affects the value of
the acceptance and induces bin-to-bin migrations in the differential
cross sections. The effect on the acceptance is relevant only in
kinematic regions near the photon $\pt$ thresholds and results in an
uncertainty of 40\% in the most affected region, the lowest values of
$m_{\Pgg\Pgg}$.  The uncertainty from the bin-to-bin migration is
about 1\%.

The systematic uncertainty on the measured photon identification
efficiency ($\varepsilon^{\text{data}}$ in Eqn.~(\ref{eq:acc})) is
estimated by applying the tag-and-probe and random-cone methods on
simulated events.
The difference between the efficiency value obtained with these methods and that
given by the fraction of simulated events passing the identification
criteria is taken as the systematic uncertainty. The uncertainty from the
acceptance and efficiency correction factor $\mathcal{C}$ is
taken as the quadratic sum of the statistical uncertainties on the
different factors of Eqn.~(\ref{eq:acc}) and the systematic uncertainty
mentioned above. The systematic and statistical uncertainties on
 $\varepsilon^{\text{data}}$ total $1.9\%$ for diphotons in the barrel
and $3.3\%$ for all diphotons.

The systematic uncertainties on the signal and background isolation
probability distributions $f(\mathcal{I})$ are estimated with Monte
Carlo pseudo-experiments in which $f(\mathcal{I})$ is varied.  The
variations correspond to the differences between the shapes of the
nominal and validation distributions observed in the validation of the
random-cone and impinging-track methods (Section
\ref{sec:extraction}). In the first bin of the probability density
functions, they are of the order of $\pm 0.01$ for the signal, and
range from $\pm 0.03$ to $\pm 0.05$ for the background.  The
uncertainty on $f(\mathcal{I})$ from its dependence on the
distribution of photon transverse energy $\ET$, photon pseudorapidity
$\eta$, and number of vertices $n_{\text{vtx}}$ is estimated from the
change in $f(\mathcal{I})$ when using the $\ET$, $\eta$, and
$n_{\text{vtx}}$ distributions from the diphoton simulation instead of
those from the diphoton event candidates in data.  This contribution to the
uncertainty is negligible.  The overall systematic uncertainty from
the $f(\mathcal{I})$ distributions on the integrated cross section is
about $8\%$, and varies from 4 to 27\% on the differential cross
sections, depending on the bin and the subsample.

A 4\% uncertainty is assigned to the integrated
luminosity~\cite{lumi}. The various contributions to the systematic
uncertainties are summarised in Table~\ref{tab:sys}.

\begin{table}[h!]
  \centering
  \caption{Contributions to the systematic uncertainties on the
    measured differential cross sections for two pseudorapidity ranges.
    The uncertainties are computed for each bin of Figs.~\ref{fig:xsec_m}
    to~\ref{fig:xsec_costhetas_eb}. The values listed below are averages.
  }

  \vspace{0.5cm}
  \begin{tabular}{l|c|c}
    Uncertainty source      & \multicolumn{1}{ c|}{$|\eta| < 1.44$} & \multicolumn{1}{c}{$|\eta| < 2.5$} \\[3pt]
    \hline
    Energy scale                    &         &         \\[-2pt]
    on acceptance                   & $1.5\%$ & $2\%$   \\[3pt]
    Energy scale                    &         &         \\[-2pt]
    on bin-to-bin migration         & $1\%$   & $1.5\%$ \\[3pt]
    Signal and background            &         &         \\[-2pt]
    distributions for $f(\mathcal{I})$   & $7\%$   & $9\%$   \\[3pt]
    Acceptance and efficiency        &         &         \\[-2pt]
    correction factor $\mathcal{C}$ & $2\%$   & $3\%$   \\[3pt]
    Luminosity                      & $4\%$ & $4\%$ \\[3pt]
    \hline
    Total                           & $8\%$   & $11\%$ \\[3pt]
    \hline
  \end{tabular}
  \label{tab:sys}
\end{table}

\section{Theoretical Predictions}
\label{sec:theory}

This section introduces the theoretical calculations whose predictions
are compared to the experimental data in Section~\ref{sec:results}.
The leading contributions to the production of pairs of prompt photons
in $\Pp\Pp$ collisions are the quark-antiquark annihilation ($\Pq\Paq
\rightarrow \Pgg\Pgg$), gluon fusion ($\Glu\Glu \rightarrow
\Pgg\Pgg$), and gluon-(anti)quark scattering ($\Pq\Glu \rightarrow
\Pgg\Pgg \Pq$) processes. One or both photons come either directly
from the hard process or from parton fragmentation, in which a cascade
of successive collinear splittings yields a radiated photon.
Contributions from the quark-antiquark annihilation process and the single- and
double-fragmentation processes are calculated up to order
$\alpha_s\alpha^2$ with the \textsc{diphox~1.3.2}
program~\cite{Binoth:1999qq}. The contributions from the gluon fusion
process, including the one-loop box diagram of order
$\alpha_s^2\alpha^2$, the interference between the one- and two-loop
box diagrams, and the real emission one-loop ``pentagon''
$\Glu\Glu\rightarrow\Pgg\Pgg \Glu$, both of order
$\alpha_s^3\alpha^2$, are calculated with the \textsc{gamma2MC 1.1.1}
program~\cite{Bern:2002jx}. The fragmentation function BFG
set~II~\cite{Bourhis:1997yu} is used in the calculation.  Although
they are higher-order processes, the gluon fusion contributions are
quantitatively comparable to those from quark-antiquark annihilation
in the diphoton mass range of interest, due to the significant gluon
luminosity in this mass range at the LHC. The three theoretical
scales, renormalisation, initial factorisation, and fragmentation, are
set to the diphoton mass value.

The photons are required to be within the kinematic acceptance defined
in Section~\ref{sec:intro}.  An additional isolation requirement at
the parton level is imposed by requiring the total hadronic transverse
energy deposited in a cone of radius $R = 0.4$ centred on the photon
direction to be less than $5\GeV$.  Particles resulting from
underlying-event activity and hadronisation are not included in
partonic event generators such as \textsc{diphox} and
\textsc{Gamma2MC}. The fraction of diphotons not selected due to
underlying hadronic activity falling inside the isolation cone is
estimated using the { \PYTHIA~6.4.22}~\cite{Sjostrand:2006za} event
generator with tunes \textsc{ z2}, {\sc d6t}~\cite{Field:2009zz}, {\sc
  p0}~\cite{Skands:2010ak}, and \textsc{ dwt}~\cite{Field:2009zz}. A
factor of $0.95 \pm 0.04$ is applied to the parton-level cross section
to correct for this effect.

The uncertainties associated with parton
distribution functions and the strong coupling constant
$\alpha_s$ are determined according to the \textsc{pdf4lhc}
recommendations~\cite{Botje:2011sn}. The diphoton cross section is computed
with three different PDF sets (\textsc{ct10}~\cite{Lai:2010vv}, \textsc{
  mstw08}~\cite{Martin:2009iq}, and \textsc{nnpdf2.1}~\cite{Ball:2011mu}),
taking into account their associated uncertainties and the
uncertainties on $\alpha_s$.  The respective preferred $\alpha_s$
central value of each PDF set is used, and $\alpha_s$ is varied by
 $\pm0.012$.  The value for the cross section is taken as the midpoint
of the envelope of the three results, including the uncertainties (68\%
confidence level envelope). The uncertainty on the cross section is
taken to be the half-width of the envelope.

The theoretical scale uncertainties are estimated by varying the
renormalisation, initial factorisation, and fragmentation scales by
factors of $1/2$ and 2, keeping the ratio between any two scales less than
2 (for example the combination $0.5\,m_{\Pgg\Pgg}$,
$2\,m_{\Pgg\Pgg}$, $m_{\Pgg\Pgg}$ is not considered). The
uncertainty is taken to be the maximum difference in the resulting
cross sections.

\section{Results}
\label{sec:results}

The integrated diphoton cross sections obtained for the acceptances
defined in Section~\ref{sec:intro} are
\begin{equation*}
  \label{equ:totxxsec}
  \begin{array}{rcl}
    \sigma(pp\rightarrow\Pgg\Pgg)|_{|\eta|< 1.44} &=& 31.0 \ \ \pm 1.8
    \  \stat \ \ \mbox{ }^{ +2.0}_{ -2.1} \  \syst \ \ \pm 1.2 \  (\text{lumi.}) \ \ \ \text{pb}~,\\
    ~\\
    \sigma(pp\rightarrow\Pgg\Pgg)|_{|\eta|< 2.50} &=& 62.4 \ \ \pm 3.6
    \ \stat \ \ \mbox{ }^{ +5.3}_{ -5.8} \  \syst \ \ \pm 2.5 \  (\text{lumi.}) \ \ \ \text{pb}.\\
  \end{array}
\end{equation*}

The theoretical calculation described in the previous section predicts
\begin{equation*}
  \label{equ:tottxsec}
  \begin{array}{rcl}
    \sigma(pp\rightarrow\Pgg\Pgg)|_{|\eta| < 1.44} & = & 27.3 \ \ \mbox{ }^{ +3.0}_{ -2.2} \ \ \text{(scales)} \ \ \pm 1.1 \ \ \text{(PDF)} \ \ \text{pb}~,\\
    ~\\
    \sigma(pp\rightarrow\Pgg\Pgg)|_{|\eta| < 2.50} & = & 52.7 \ \ \mbox{ }^{ +5.8}_{ -4.2} \ \ \text{(scales)} \ \ \pm 2.0 \ \ \text{(PDF)} \ \ \text{pb}.\\
  \end{array}
\end{equation*}
The integrated cross sections obtained from the calculation are
consistent with the measurements within the experimental and
theoretical uncertainties.

\begin{figure}[p]
  \centering
    \includegraphics[width=0.45\textwidth] {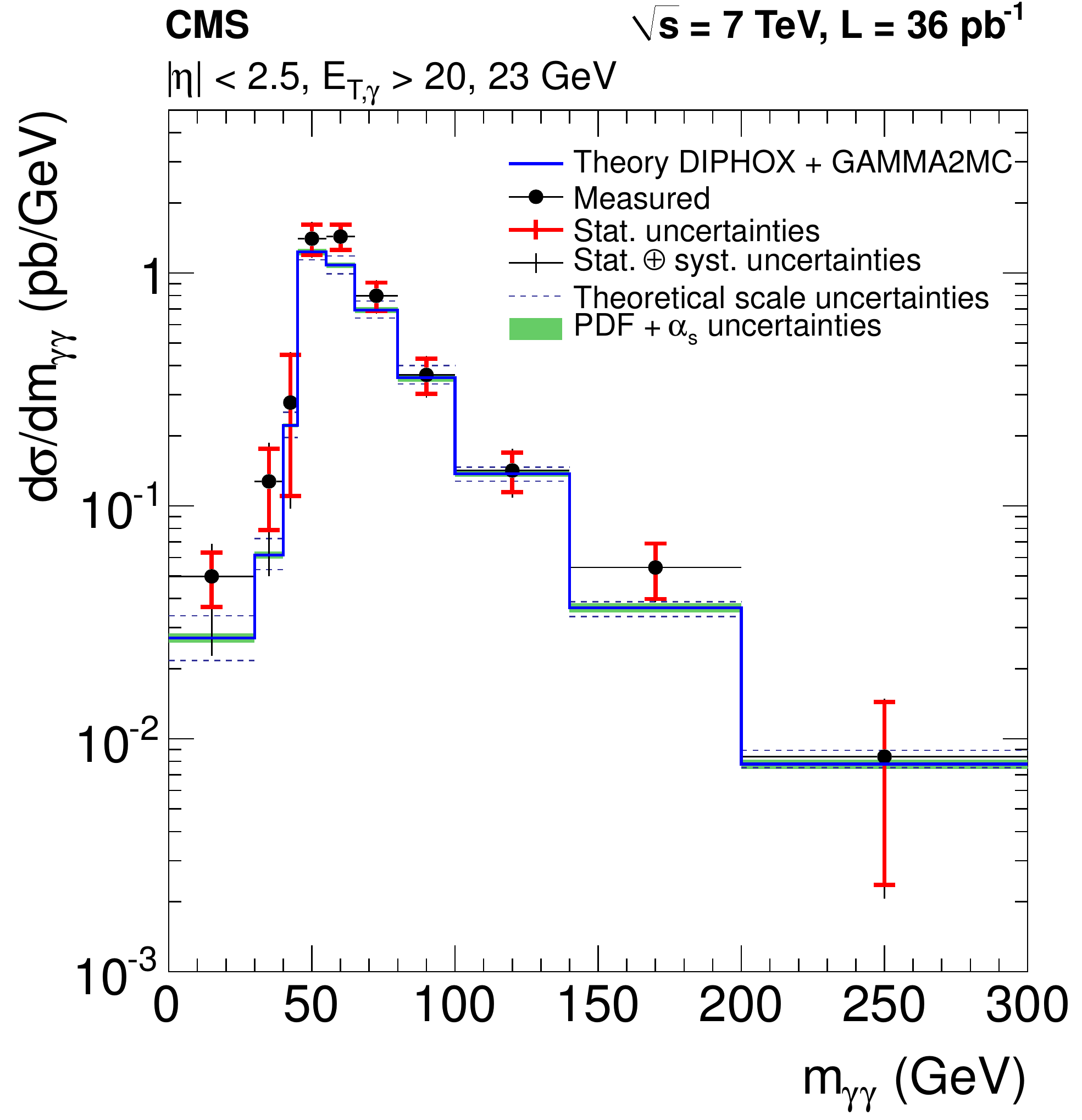}
    \includegraphics[width=0.45\textwidth] {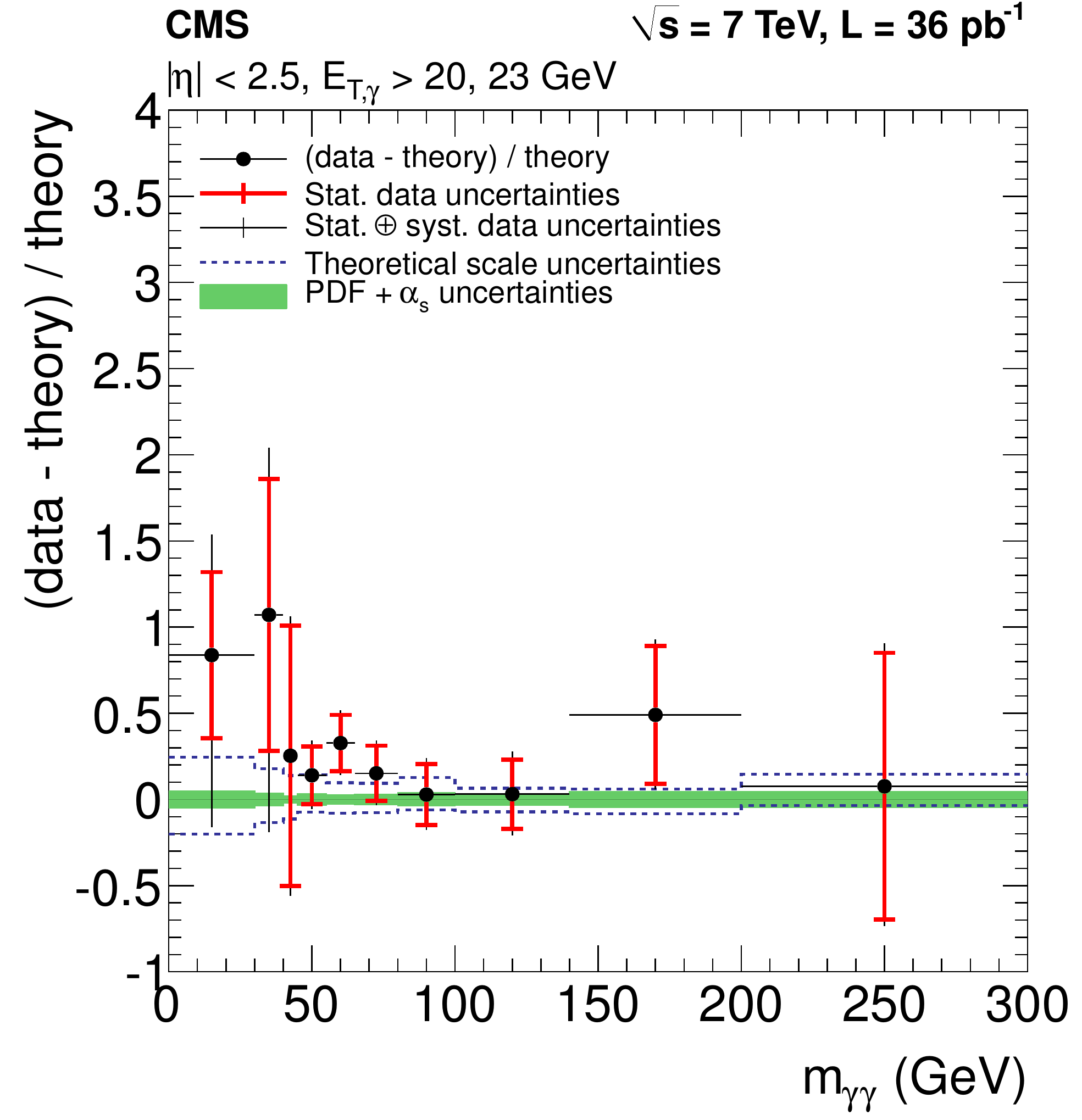}
  \caption{\label{fig:xsec_m}
    (Left) Diphoton differential cross section as a function of the photon
    pair invariant mass $m_{\Pgg\Pgg}$ from data (points) and from theory
    (solid line) for the photon pseudorapidity range $|\eta| < 2.5$.
    (Right) The difference between the measured and theoretically predicted
    diphoton cross sections, divided by the theory prediction, as a function of
    $m_{\Pgg\Pgg}$.
    In both plots, the inner and outer error bars on each point show the
    statistical and total experimental uncertainties. The 4\% uncertainty on the integrated luminosity
      is not included in the error bars.
      The dotted line and shaded region represent the systematic uncertainties
    on the theoretical prediction from the theoretical scales and the PDFs, respectively.}
\end{figure}

\begin{figure}[p]
  \centering
    \includegraphics[width=0.45\textwidth] {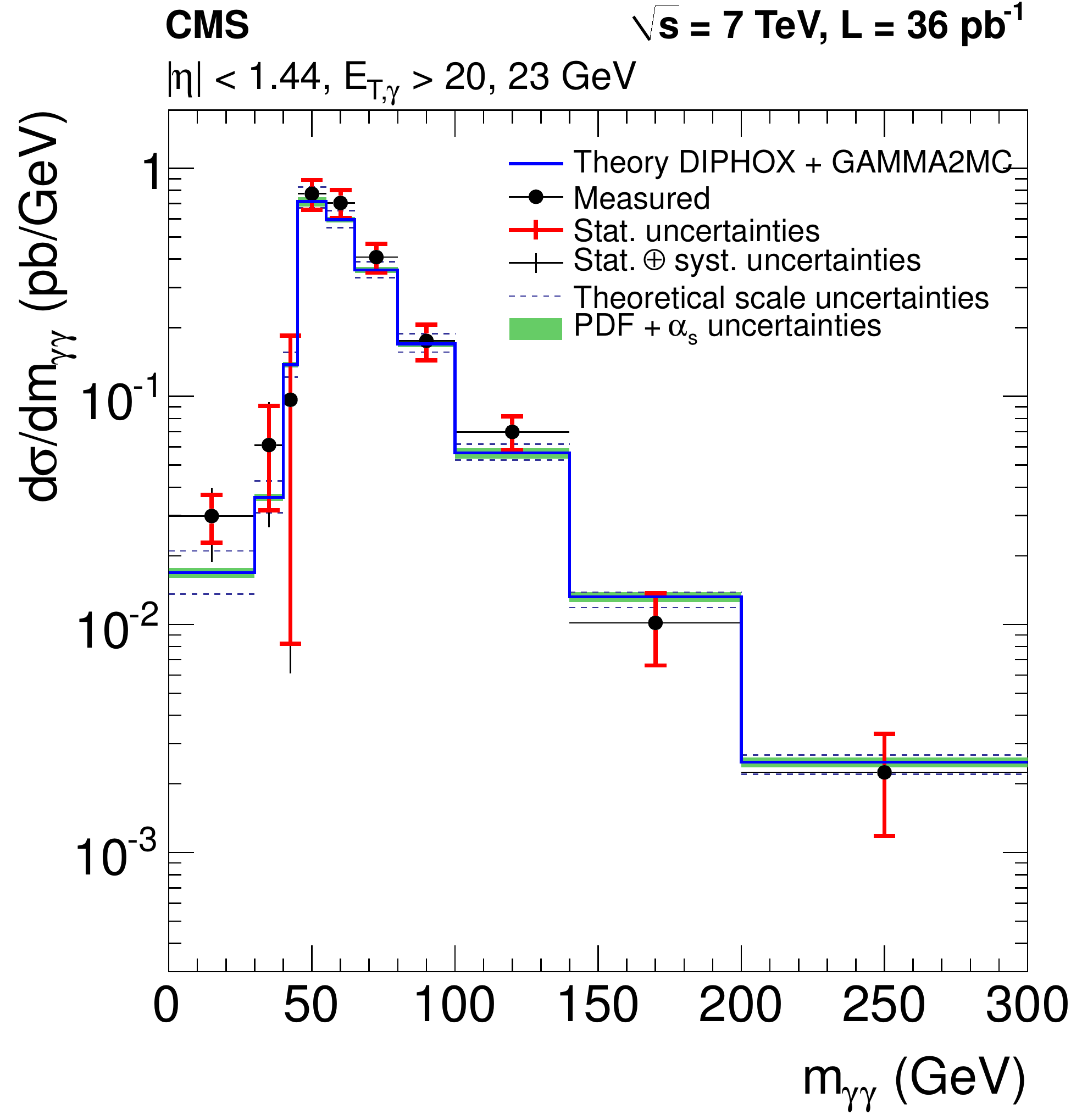}
    \includegraphics[width=0.45\textwidth] {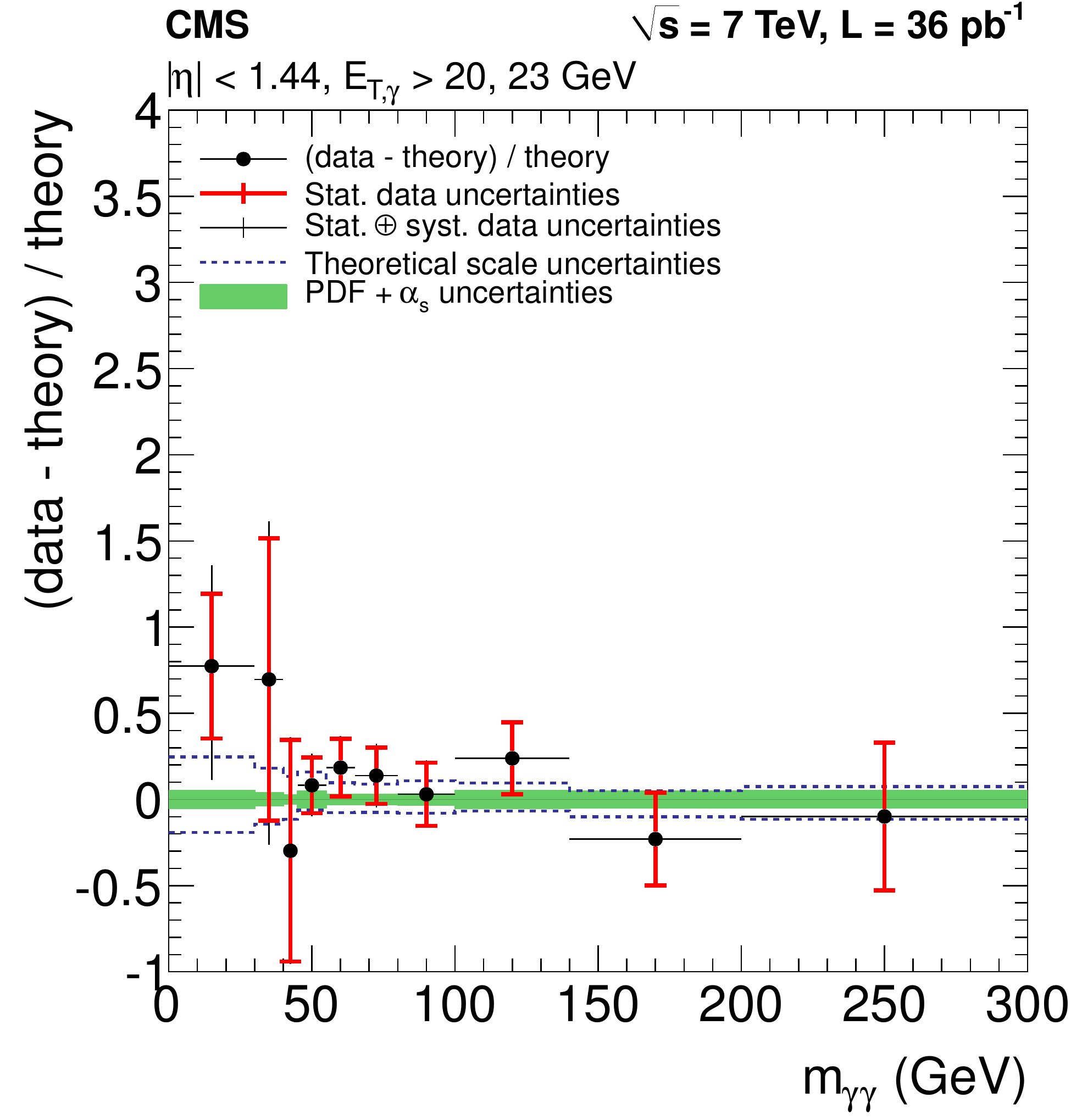}
    \caption{(Left) Diphoton differential cross section as a function
      of the photon pair invariant mass $m_{\Pgg\Pgg}$ from data
      (points) and from theory (solid line) for the photon
      pseudorapidity range $|\eta| < 1.44$.  (Right) The difference
      between the measured and theoretically predicted diphoton cross
      sections, divided by the theory prediction, as a function of
      $m_{\Pgg\Pgg}$.  In both plots, the inner and outer error bars
      on each point show the statistical and total experimental
      uncertainties. The 4\% uncertainty on the integrated luminosity
      is not included in the error bars. The dotted line and shaded
      region represent the systematic uncertainties on the theoretical
      prediction from the theoretical scales and the PDFs,
      respectively.}
  \label{fig:xsec_m_eb}
\end{figure}

\begin{figure}[p]
  \centering
    \includegraphics[width=0.45\textwidth] {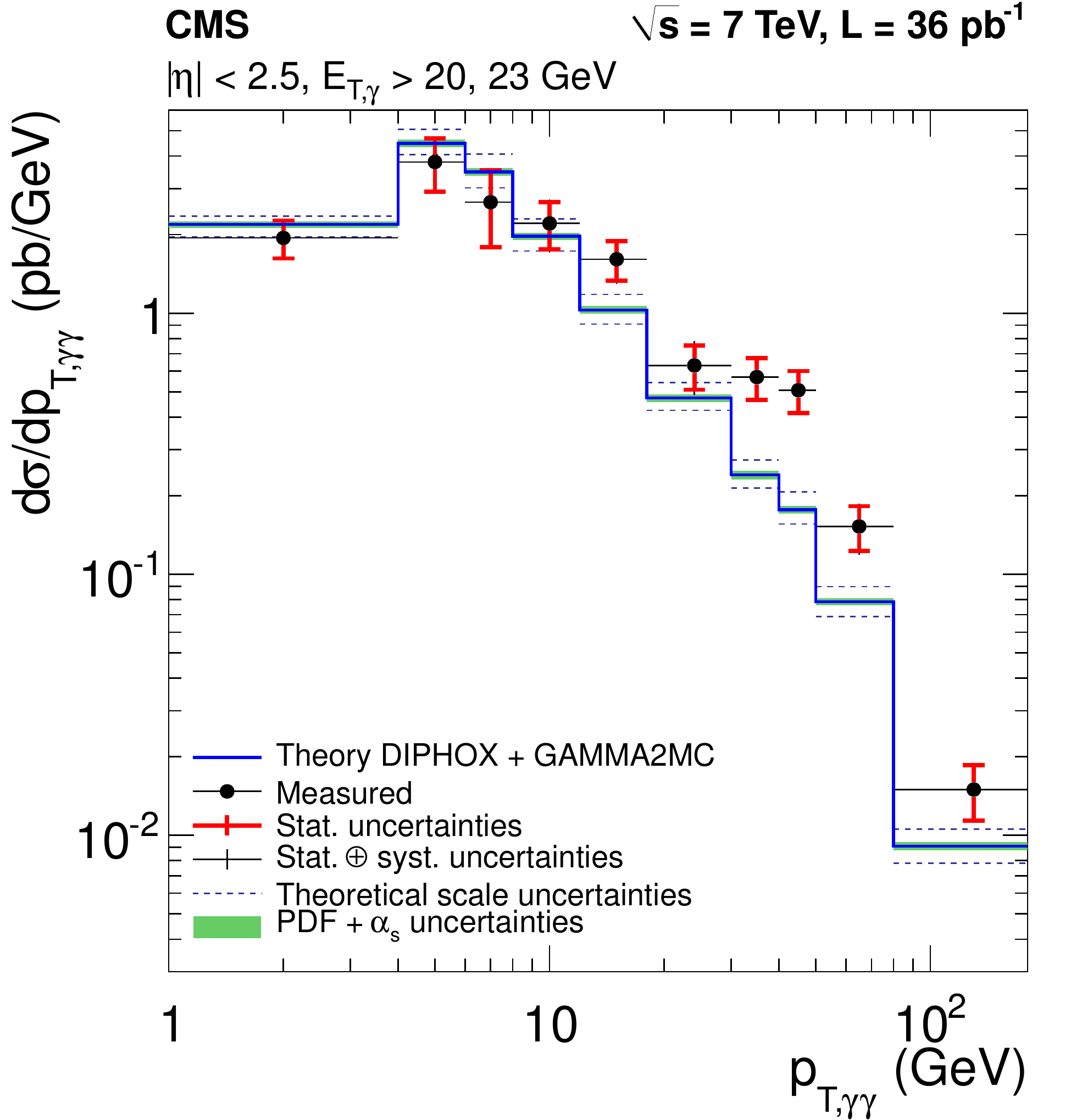}
    \includegraphics[width=0.45\textwidth] {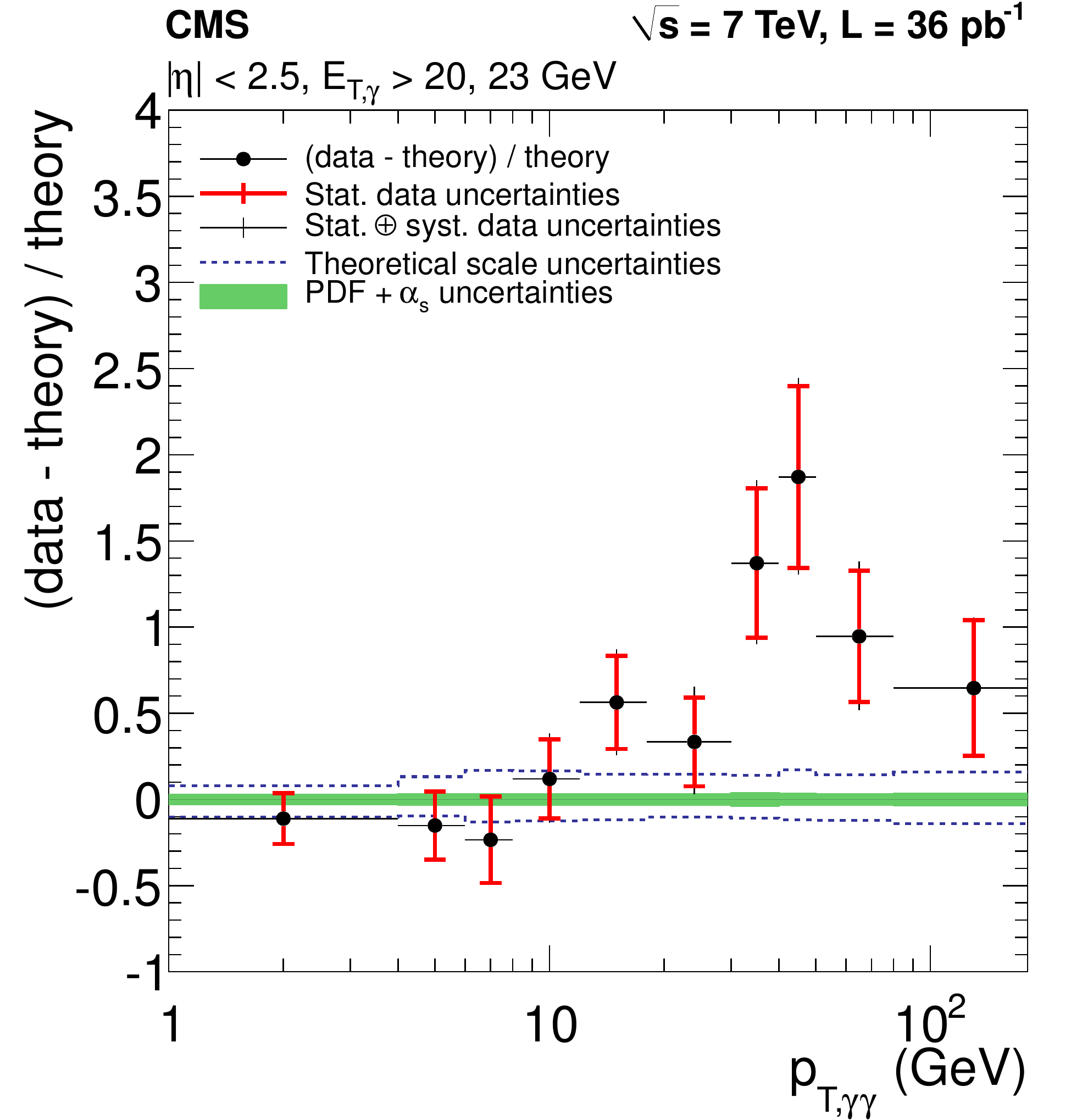}
    \caption{(Left) Diphoton differential cross section as a function
      of the photon pair transverse momentum $p_{T,\Pgg\Pgg}$ from
      data (points) and from theory (solid line) for the photon
      pseudorapidity range $|\eta| < 2.5$.  (Right) The difference
      between the measured and theoretically predicted diphoton cross
      sections, divided by the theory prediction, as a function of
      $p_{T,\Pgg\Pgg}$.  In both plots, the inner and outer error bars
      on each point show the statistical and total experimental
      uncertainties. The 4\% uncertainty on the integrated luminosity
      is not included in the error bars. The dotted line and shaded
      region represent the systematic uncertainties on the theoretical
      prediction from the theoretical scales and the PDFs,
      respectively.}
  \label{fig:xsec_pt}
\end{figure}

\begin{figure}[p]
  \centering
    \includegraphics[width=0.45\textwidth] {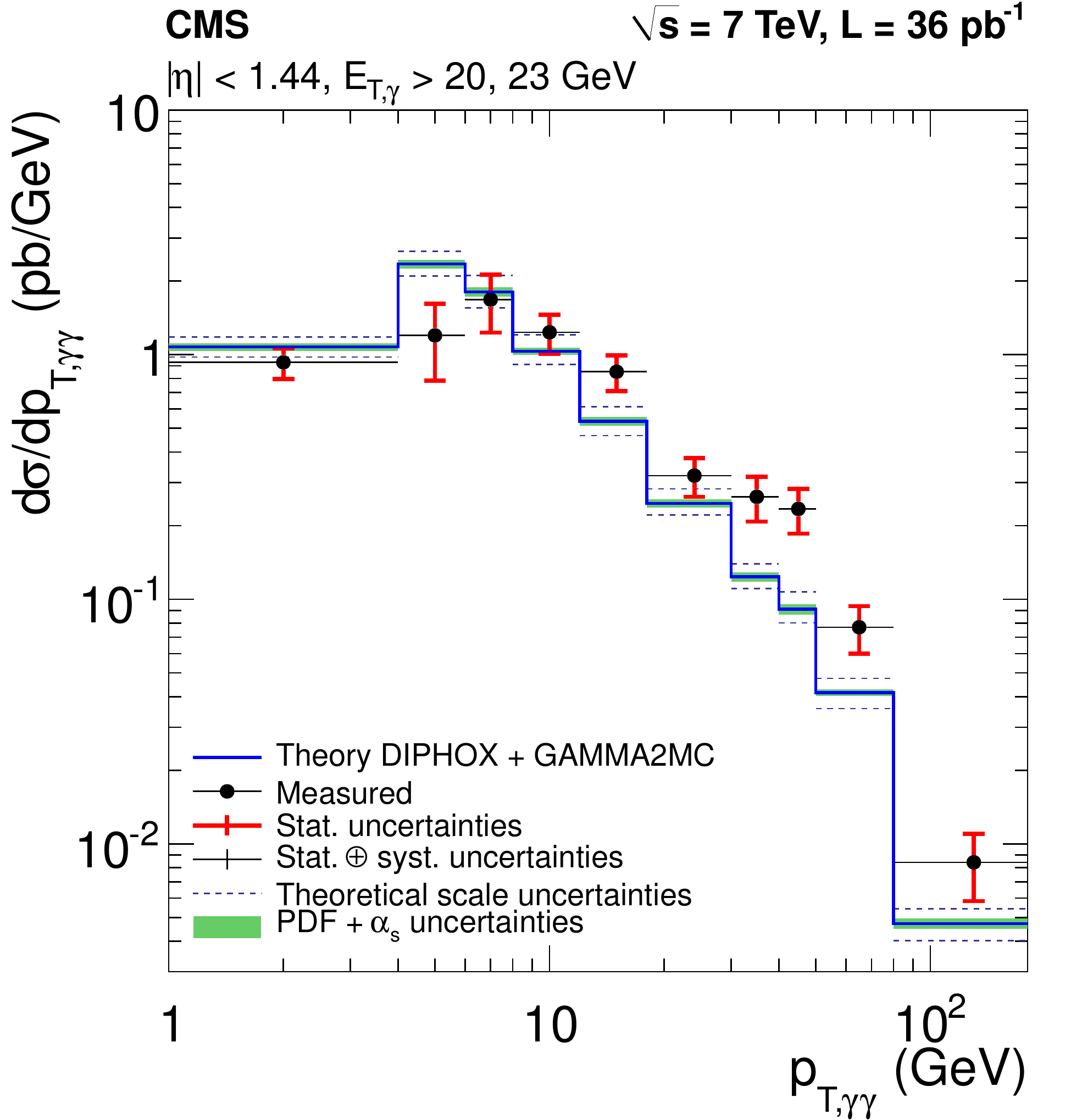}
    \includegraphics[width=0.45\textwidth] {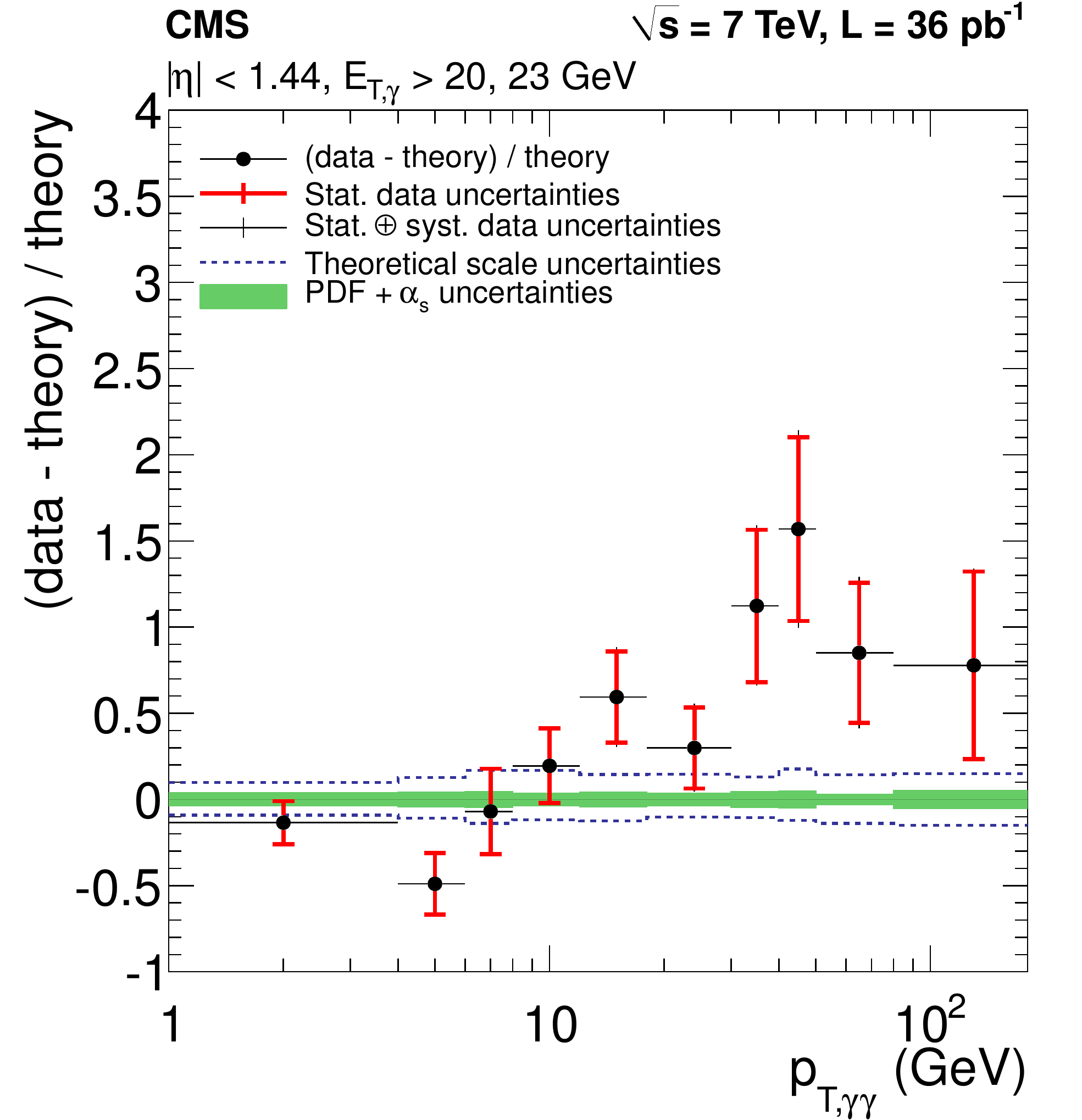}
    \caption{(Left) Diphoton differential cross section as a function
      of the photon pair transverse momentum $p_{T,\Pgg\Pgg}$ from
      data (points) and from theory (solid line) for the photon
      pseudorapidity range $|\eta| < 1.44$.  (Right) The difference
      between the measured and theoretically predicted diphoton cross
      sections, divided by the theory prediction, as a function of
      $p_{T,\Pgg\Pgg}$.  In both plots, the inner and outer error bars
      on each point show the statistical and total experimental
      uncertainties. The 4\% uncertainty on the integrated luminosity
      is not included in the error bars. The dotted line and shaded
      region represent the systematic uncertainties on the theoretical
      prediction from the theoretical scales and the PDFs,
      respectively.}
  \label{fig:xsec_pt_eb}
\end{figure}

\begin{figure}[p]
  \centering
    \includegraphics[width=0.45\textwidth] {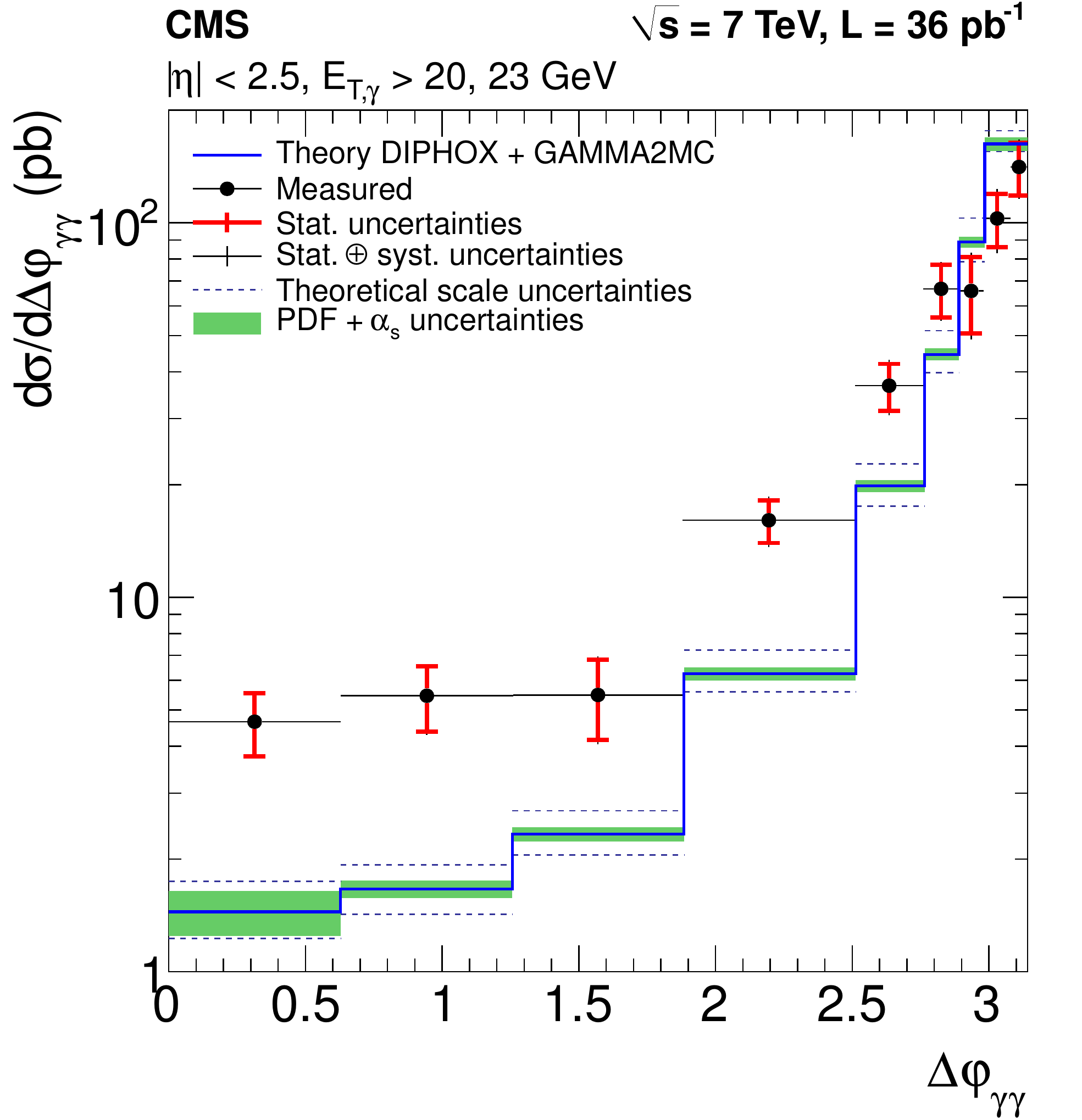}
    \includegraphics[width=0.45\textwidth] {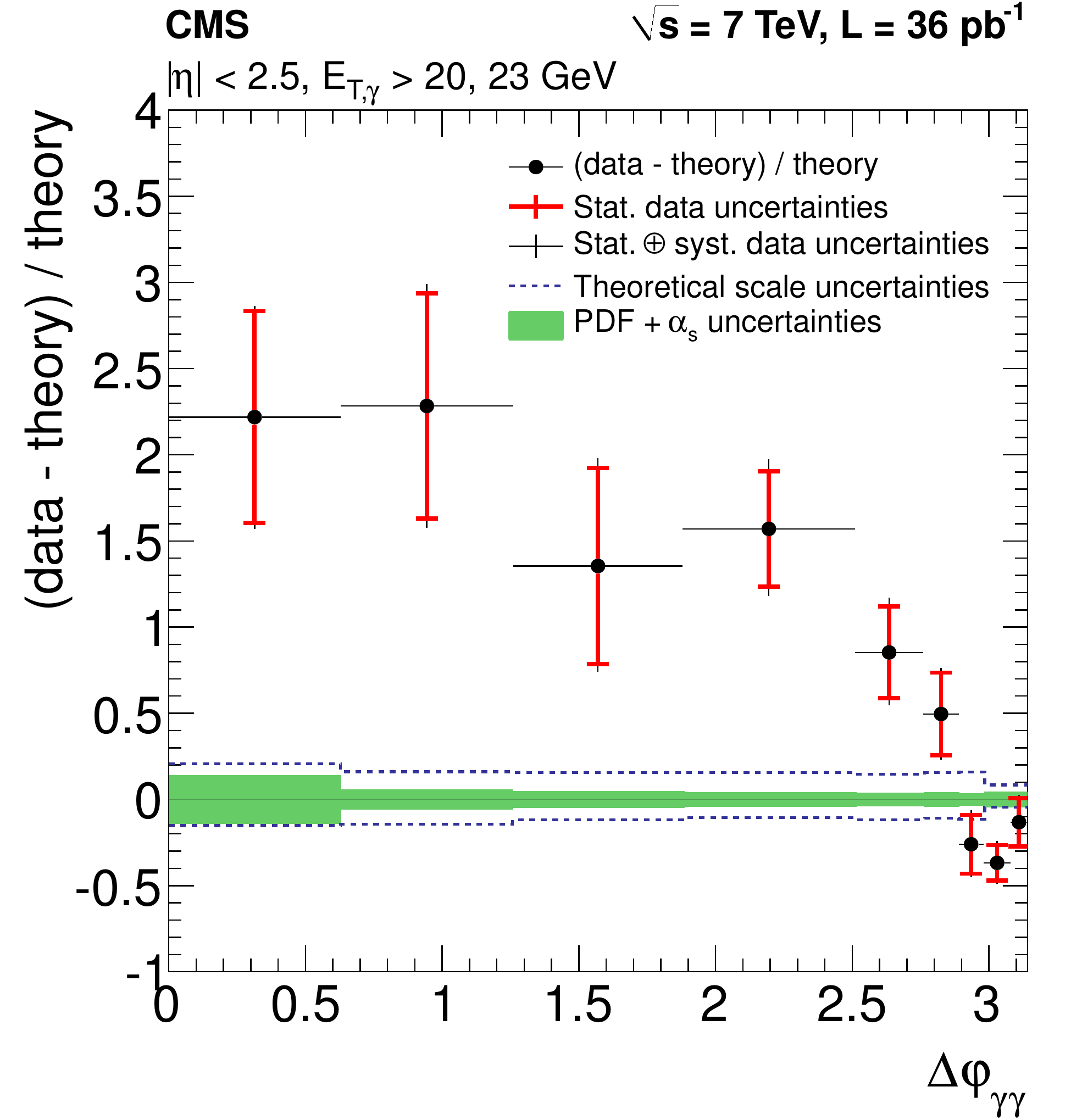}
    \caption{\label{fig:xsec_dphi} (Left) Diphoton differential cross
      section as a function of the azimuthal angle between the two
      photons, $\Delta\varphi_{\Pgg\Pgg}$, from data (points) and from
      theory (solid line) for the photon pseudorapidity range $|\eta|
      <2.5$.  (Right) The difference between the measured and
      theoretically predicted diphoton cross sections, divided by the
      theory prediction, as a function of $\Delta\varphi_{\Pgg\Pgg}$.
      In both plots, the inner and outer error bars on each point show
      the statistical and total experimental uncertainties. The 4\%
      uncertainty on the integrated luminosity is not included in the
      error bars. The dotted line and shaded region represent the
      systematic uncertainties on the theoretical prediction from the
      theoretical scales and the PDFs, respectively.}
\end{figure}

\begin{figure}[p]
  \centering
    \includegraphics[width=0.45\textwidth] {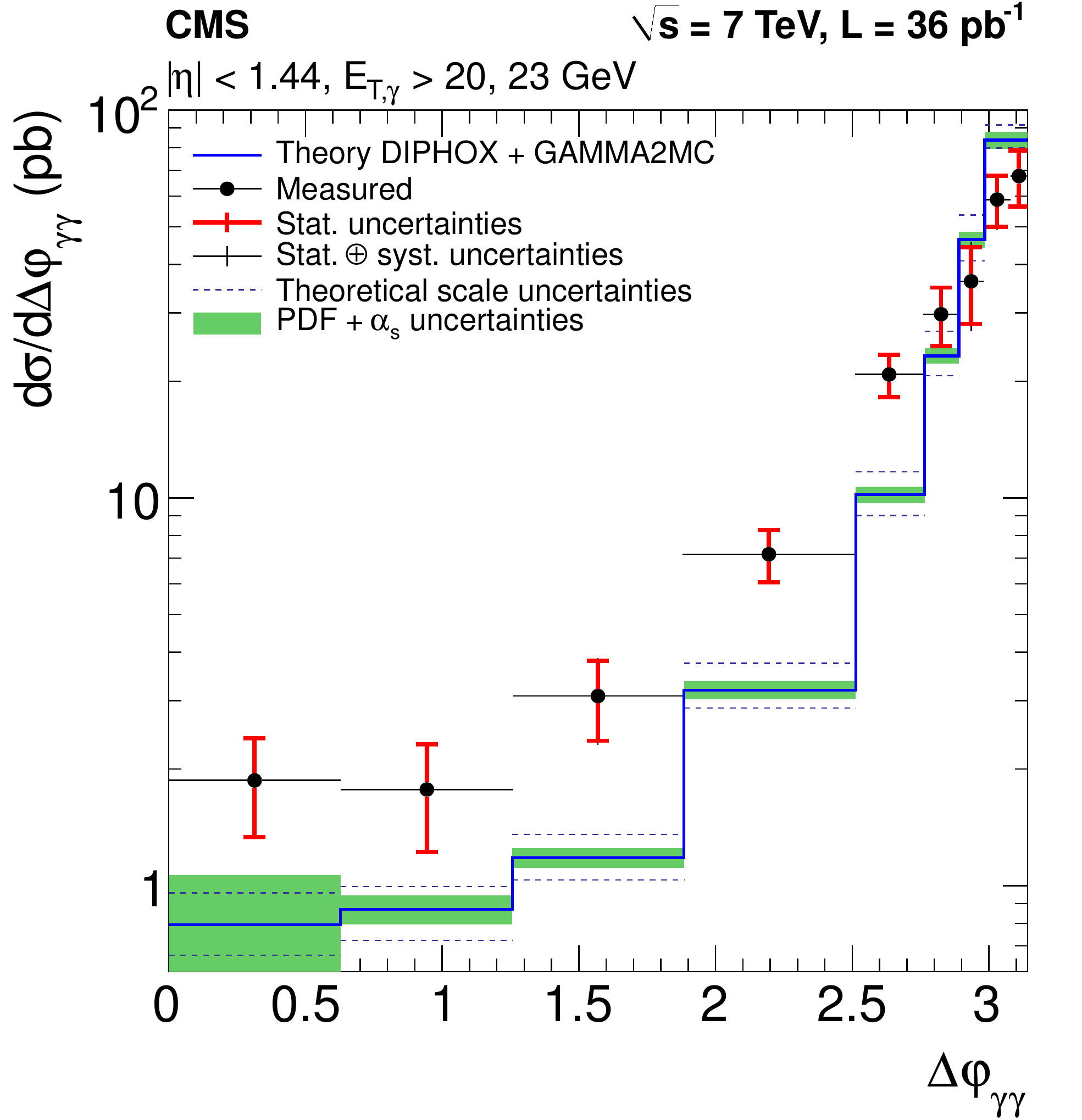}
    \includegraphics[width=0.45\textwidth] {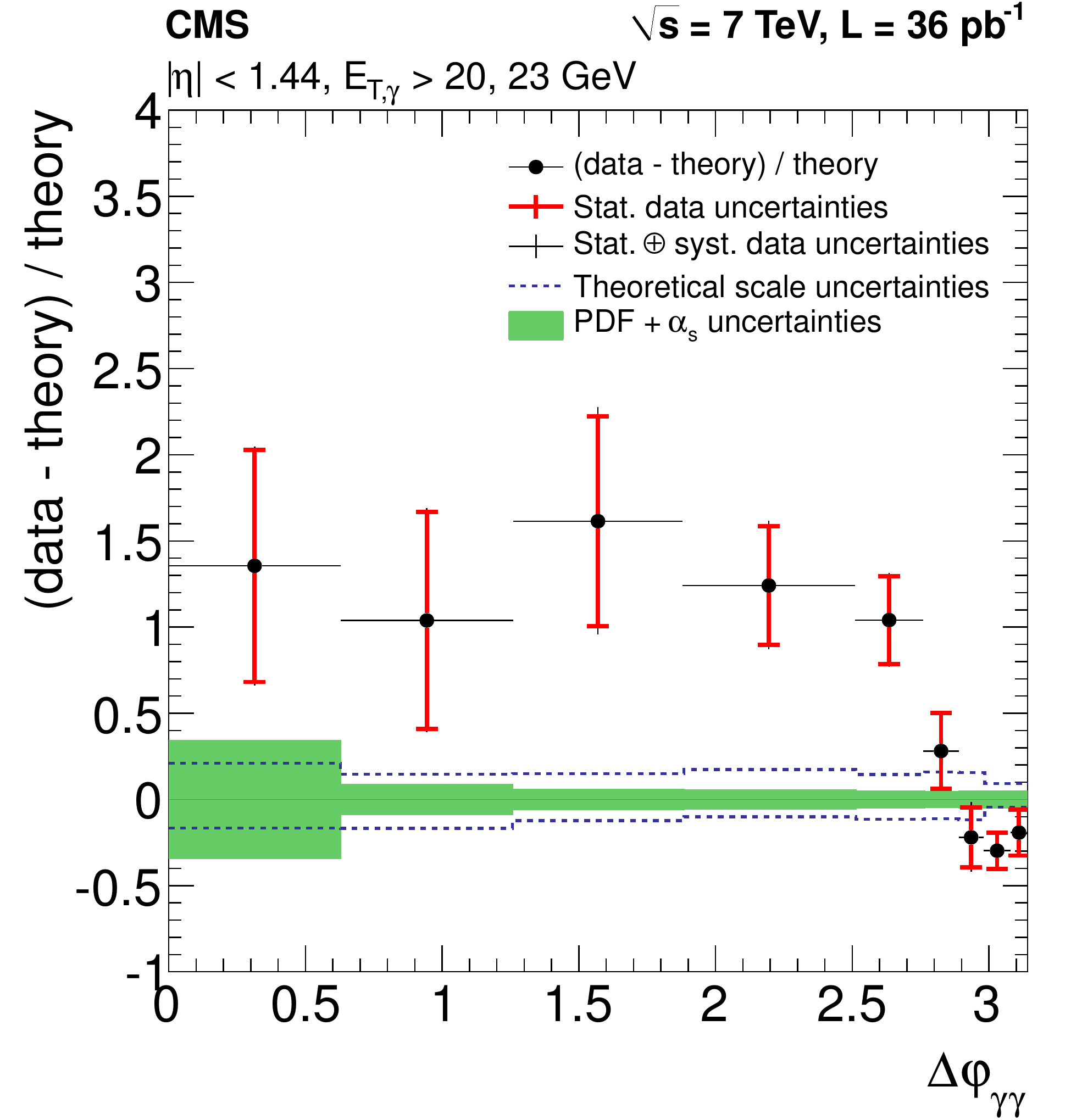}
    \caption{(Left) Diphoton differential cross section as a function
      of the azimuthal angle between the two photons, $\Delta\varphi_{\Pgg\Pgg}$,
      from data (points) and from theory
      (solid line) for the photon pseudorapidity range $|\eta| <1.44$.
      (Right) The difference between the measured and theoretically
      predicted diphoton cross sections, divided by the theory
      prediction, as a function of $\Delta\varphi_{\Pgg\Pgg}$.  In
      both plots, the inner and outer error bars on each point show
      the statistical and total experimental uncertainties. The 4\%
      uncertainty on the integrated luminosity is not included in the
      error bars. The dotted line and shaded region represent the
      systematic uncertainties on the theoretical prediction from the
      theoretical scales and the PDFs, respectively.}
  \label{fig:xsec_dphi_eb}
\end{figure}

\begin{figure}[p]
  \centering
    \includegraphics[width=0.45\textwidth] {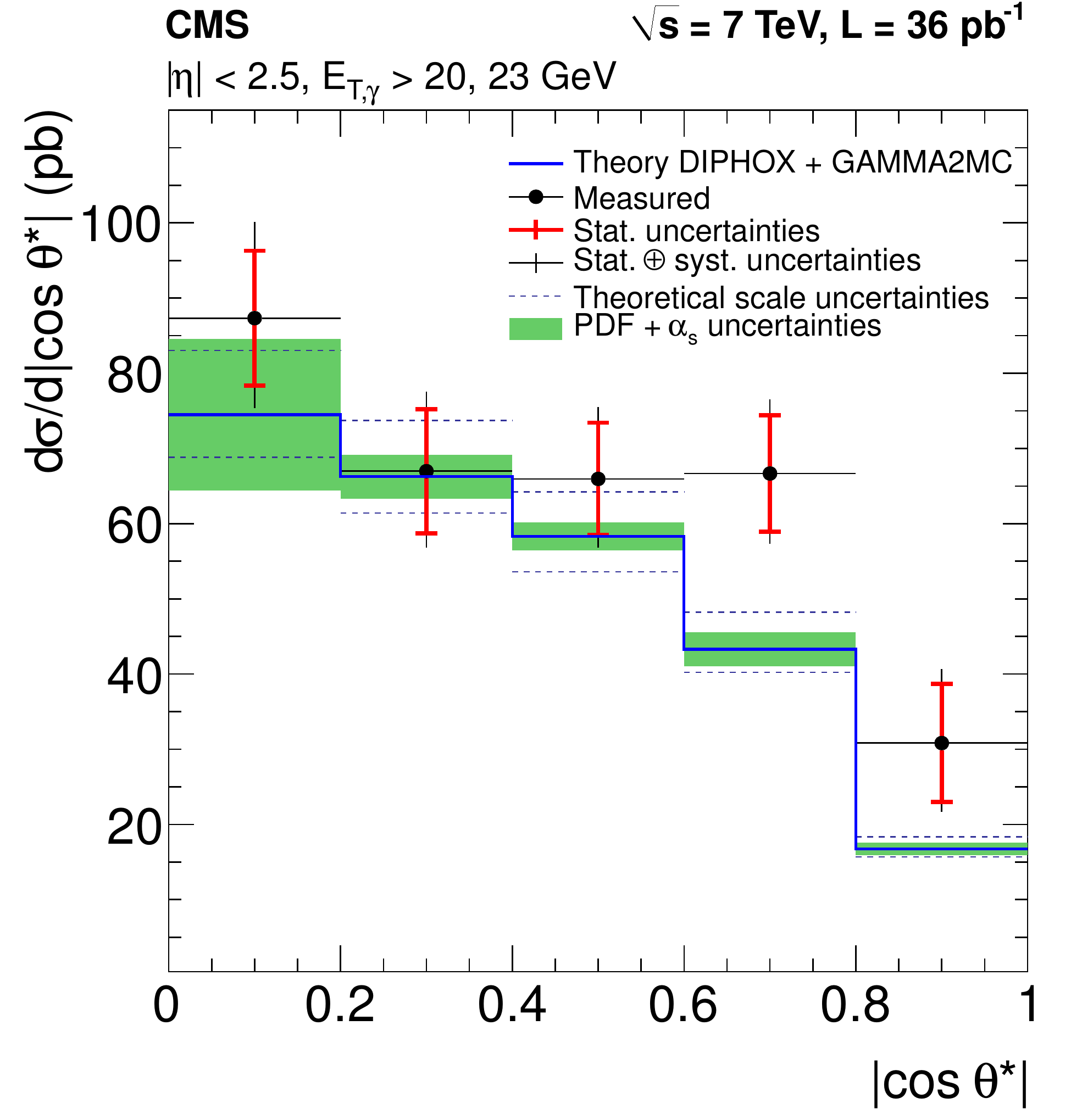}
    \includegraphics[width=0.45\textwidth] {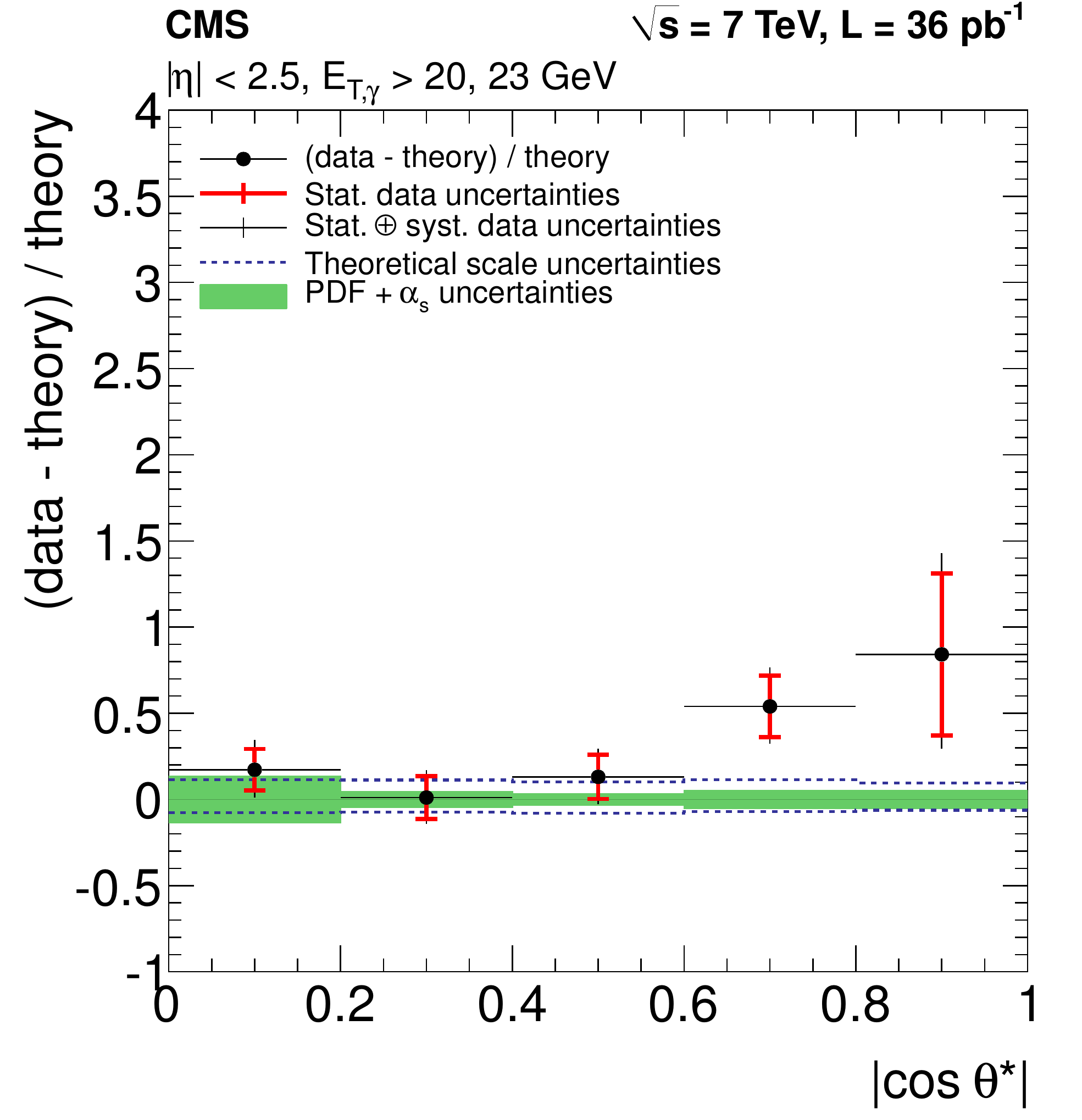}
    \caption{(Left) Diphoton differential cross section as a function
      of $|{\cos\theta^*}|$ from data (points) and from theory (solid
      line) for the photon pseudorapidity range $|\eta| <2.5$.
      (Right) The difference between the measured and theoretically
      predicted diphoton cross sections, divided by the theory
      prediction, as a function of $|{\cos\theta^*}|$.  In both plots,
      the inner and outer error bars on each point show the
      statistical and total experimental uncertainties. The 4\%
      uncertainty on the integrated luminosity is not included in the
      error bars. The dotted line and shaded region represent the
      systematic uncertainties on the theoretical prediction from the
      theoretical scales and the PDFs, respectively.}
  \label{fig:xsec_costhetas}
\end{figure}

\begin{figure}[p]
  \centering
    \includegraphics[width=0.45\textwidth] {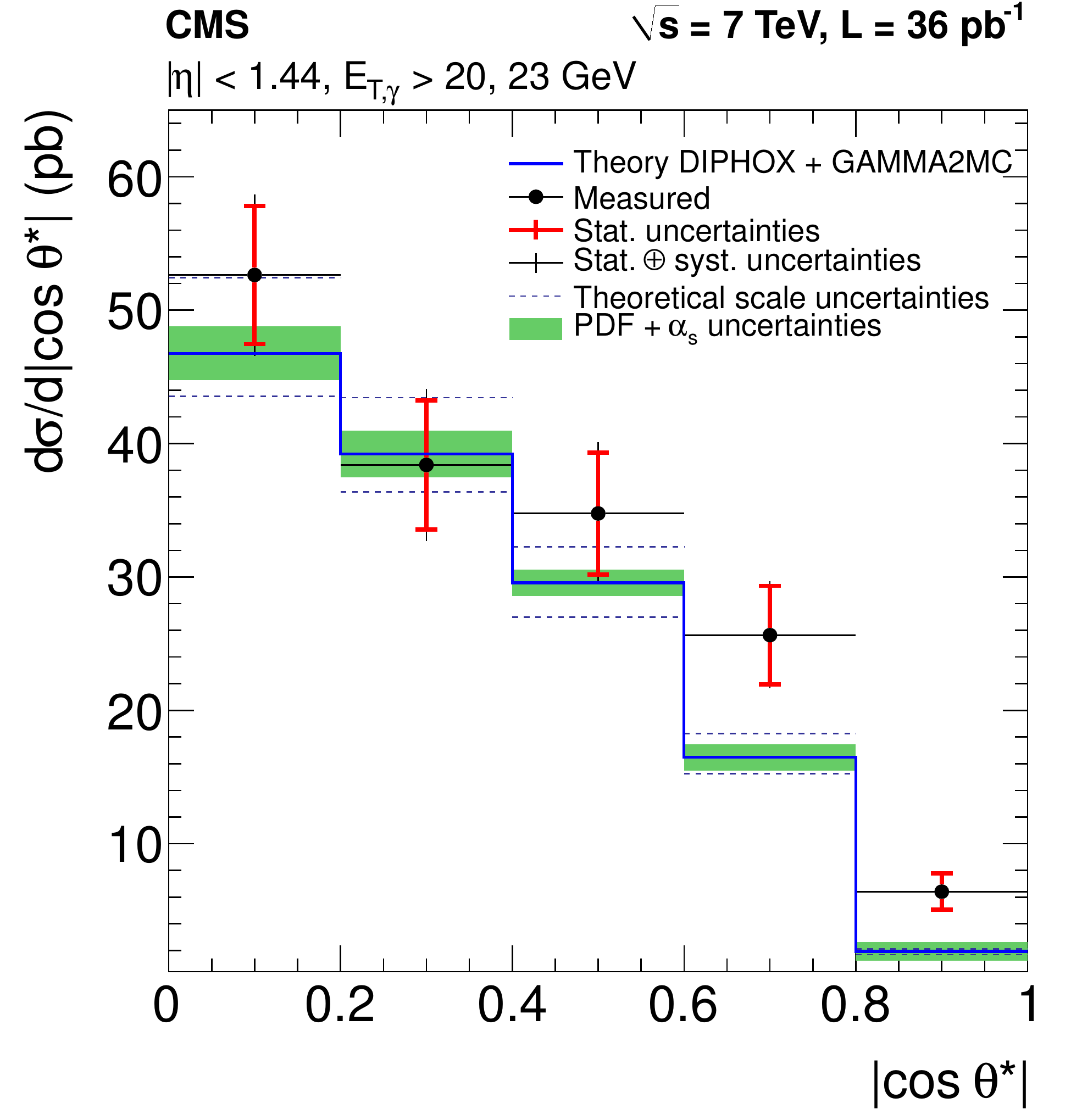}
    \includegraphics[width=0.45\textwidth] {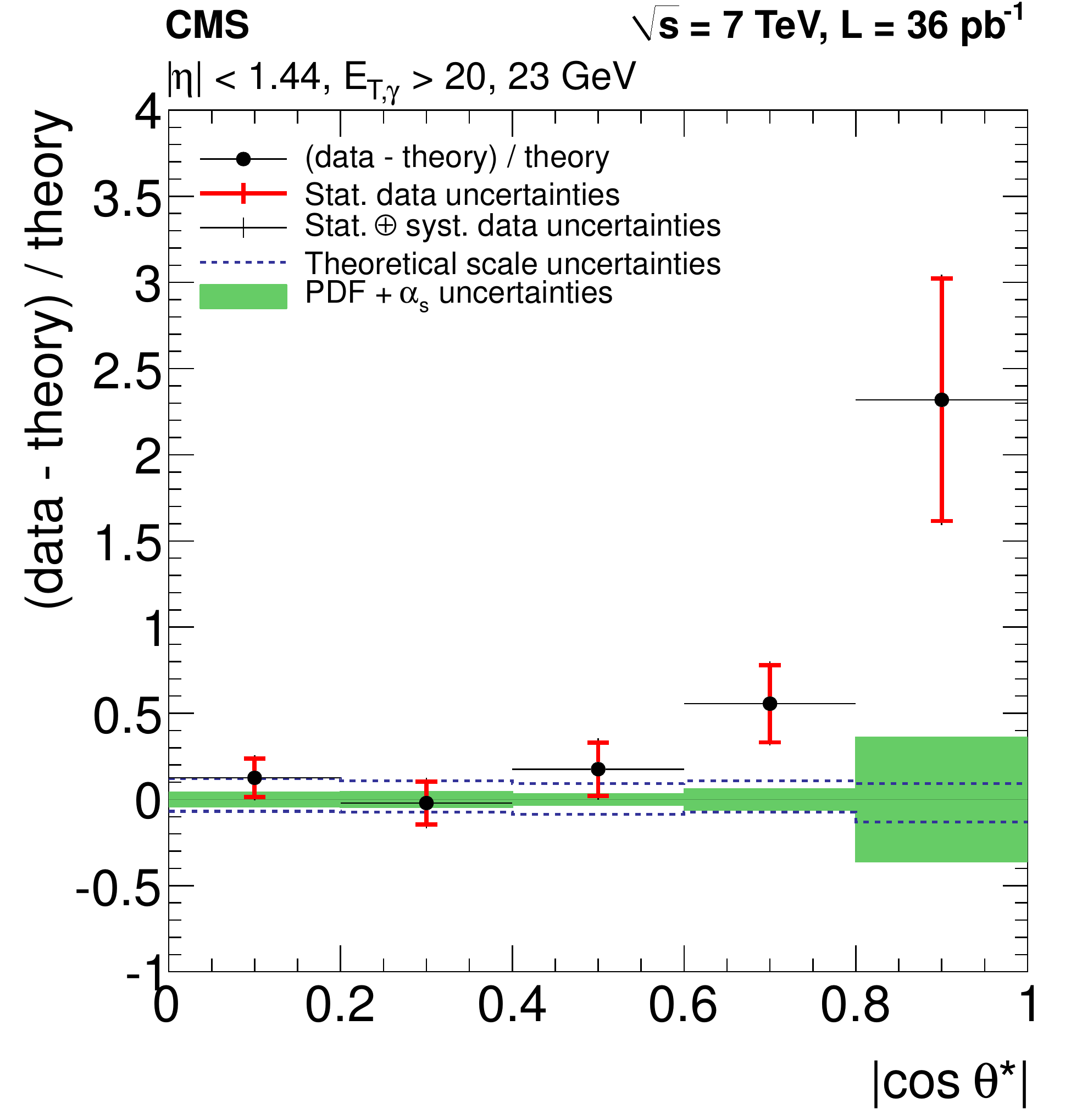}
    \caption{(Left) Diphoton differential cross section as a function
      of $|{\cos\theta^*}|$ from data (points) and from theory (solid
      line) for the photon pseudorapidity range $|\eta| <1.44$.
      (Right) The difference between the measured and theoretically
      predicted diphoton cross sections, divided by the theory
      prediction, as a function of $|{\cos\theta^*}|$.  In both plots,
      the inner and outer error bars on each point show the
      statistical and total experimental uncertainties. The 4\%
      uncertainty on the integrated luminosity is not included in the
      error bars. The dotted line and shaded region represent the
      systematic uncertainties on the theoretical prediction from the
      theoretical scales and the PDFs, respectively.}
  \label{fig:xsec_costhetas_eb}
\end{figure}

The differential cross-section measurements as functions of
$m_{\Pgg\Pgg}$, $\Delta\varphi_{\Pgg\Pgg}$, $p_{T,\Pgg\Pgg}$, and
$|{\cos\theta^*}|$ for the two pseudorapidity ranges are shown, along
with the theoretical predictions, in Figs.~\ref{fig:xsec_m} to
\ref{fig:xsec_costhetas_eb}. The $4\%$ uncertainty on the integrated
luminosity is not included in the error bars.  The values of
the cross sections are given in Tables \ref{tab:xsec_m} to
\ref{tab:xsec_costhetas}.  As can be seen in Figs.~\ref{fig:xsec_dphi}
and~\ref{fig:xsec_dphi_eb}, the theoretical predictions underestimate
the measured cross section for $\Delta\varphi_{\Pgg\Pgg} < 2.8$.  In
the leading-order (LO) diagrams of gluon fusion and quark-antiquark annihilation
$2\rightarrow 2$ processes, the two photons are back-to-back because
of momentum conservation. Therefore, the LO term does not contribute
to this phase space region, which thus only receives contributions
from NLO terms for both the direct and fragmentation diphoton
production processes.

The contribution for $\Delta\varphi_{\Pgg\Pgg} \lesssim
2.8$, combined with the requirements of $E_{T} > 20$ and $23\GeV$ on
the two photons, is responsible for the shoulder around $40\GeV$ in
the diphoton differental $p_{T}$ distribution of
Figs.~\ref{fig:xsec_pt} and~\ref{fig:xsec_pt_eb}.  This contribution
also populates the region below $30\GeV$ in the diphoton mass
distribution shown in Figs.~\ref{fig:xsec_m}
and~\ref{fig:xsec_m_eb}. In these two regions of the $p_{T,\Pgg\Pgg}$
and $m_{\Pgg\Pgg}$ spectra, the theoretical cross section is lower
than the measurement, consistent with the deficit for
$\Delta\varphi_{\Pgg\Pgg}\lesssim2.8$.

Comparison of the measurements of the $|{\cos\theta^{*}}|$ spectra with
theoretical predictions, shown in Figs.~\ref{fig:xsec_costhetas}
and~\ref{fig:xsec_costhetas_eb}, reveals an underestimation from the
theory at large $|{\cos\theta^{*}}|$ values, which is more significant
for the central rapidity range ($|{\eta}| < 1.44$).  Similar discrepancies
have previously been observed in diphoton production at hadron
colliders~\cite{Aaltonen:2011vi, Abazov:2010ah, atlas2011} as
discussed in Ref.~\cite{PhysRevD.63.114016}.

\begin{table}[htpb]
  \centering
  \caption{Measured diphoton differential cross section as a function of
    $m_{\Pgg\Pgg}$ for the two photon pseudorapidity ranges, with statistical $\stat$ and systematic $\syst$ uncertainties.}
  \vspace{0.5cm}
  \begin{tabular}{r@{}c@{}l|r@{}l @{\hspace{1em}} l@{\hspace{1em}}l@{\ \ }l | r@{}l @{\hspace{1em}} ll@{\ \ }l}
    \hline
    \hline\\[-.3cm]
    \multicolumn{13}{c}{\Large$d\sigma/dm_{\Pgg\Pgg}$ [$\text{pb}/\!\GeV$]}\\[3pt]
    \hline
    \multicolumn{3}{c|}{$m_{\Pgg\Pgg}$ [$\GeV$]}& \multicolumn{5}{c|}{$|\eta| < 1.44$} & \multicolumn{5}{c}{$|\eta| < 2.5$}\\[3pt] \hline
    & & &       &     &\multicolumn{1}{c}{stat.}&\multicolumn{2}{c|}{syst.}           &         &   &\multicolumn{1}{c}{stat.}   &\multicolumn{2}{c}{syst.}\\[3pt]
    $0$ &$-$&$30$  & $0$&$.0299$ &$\pm0.0071  $&$+0.0069$&$-0.0086$ & $0$&$.050$ &$\pm0.013  $&$+0.014$ & $-0.024$ \\[3pt]
    $30$&$-$&$40$  & $0$&$.061$  &$\pm0.030   $&$+0.015$ & $-0.018$ & $0$&$.127$ &$\pm0.049  $&$+0.035$ & $-0.061$ \\[3pt]
    $40$&$-$&$45$  & $0$&$.097$  &$\pm0.088   $&$+0.020$ & $-0.020$ & $0$&$.28$  &$\pm0.17   $&$+0.06$  & $-0.07$  \\[3pt]
    $45$&$-$&$55$  & $0$&$.77$   &$\pm0.12    $&$+0.06$  & $-0.05$  & $1$&$.40$  &$\pm0.20   $&$+0.14$  & $-0.12$  \\[3pt]
    $55$&$-$&$65$  & $0$&$.70$   &$\pm0.10    $&$+0.05$  & $-0.04$  & $1$&$.43$  &$\pm0.18   $&$+0.10$  & $-0.09$  \\[3pt]
    $65$&$-$&$80$  & $0$&$.408$  &$\pm0.059   $&$+0.030$ & $-0.031$ & $0$&$.80$  &$\pm0.11   $&$+0.07$  & $-0.06$  \\[3pt]
    $80$&$-$&$100$ & $0$&$.175$  &$\pm0.031   $&$+0.013$ & $-0.012$ & $0$&$.365$ &$\pm0.063  $&$+0.041$ & $-0.037$ \\[3pt]
    $100$&$-$&$140$& $0$&$.070$  &$\pm0.012   $&$+0.003$ & $-0.003$ & $0$&$.142$ &$\pm0.028  $&$+0.020$ & $-0.018$ \\[3pt]
    $140$&$-$&$200$& $0$&$.0102$ &$\pm0.0035  $&$+0.0007$& $-0.0006$& $0$&$.054$ &$\pm0.015$  &$+0.006$ & $-0.006$ \\[3pt]
    $200$&$-$&$300$ & $0$&$.0022$ &$\pm0.0011 $&$+0.0001$& $-0.0001$& $0$&$.0084$&$\pm0.0060$ &$+0.0023$& $-0.0019$ \\[3pt]
    \hline
    \hline
  \end{tabular}
  \label{tab:xsec_m}
\end{table}

\begin{table}[htpb]
    \centering
    \caption{Measured diphoton differential cross section as a function of
      $p_{T,\Pgg\Pgg}$ for the two photon pseudorapidity ranges, with statistical (stat.) and systematic (syst.) uncertainties.}
    \vspace{0.5cm}
     \begin{tabular}{r@{}c@{}l|r@{}l @{\hspace{1em}} l@{\hspace{1em}}l@{\ \ }l | r@{}l @{\hspace{1em}} ll@{\ \ }l}
        \hline
         \hline\\[-.3cm]
        \multicolumn{13}{c}{\Large$d\sigma/dp_{T,\Pgg\Pgg}$
        [$\text{pb}/\!\GeV$]}\\[3pt]
        \hline
        \multicolumn{3}{c|}{$p_{T,\Pgg\Pgg}$ [$\GeV$]}& \multicolumn{5}{c|}{$|\eta| < 1.44$} & \multicolumn{5}{c}{$|\eta| < 2.5$}\\[3pt] \hline
         & & &        &     &\multicolumn{1}{c}{stat.}&\multicolumn{2}{c|}{syst.}           &         & &\multicolumn{1}{c}{stat.}   &\multicolumn{2}{c}{syst.}\\[3pt]
$0$&$-$&$4$   & $0$&$.93$  &$\pm0.13   $&$+0.04$  & $-0.05$  & $1$&$.94$ &$\pm0.32   $&$+0.12$  & $-0.13$ \\[3pt]
$4$&$-$&$6$   & $1$&$.20$  &$\pm0.42   $&$+0.10$  & $-0.09$  & $3$&$.80$ &$\pm0.88   $&$+0.27$  & $-0.29$ \\[3pt]
$6$&$-$&$8$   & $1$&$.68$  &$\pm0.45   $&$+0.12$  & $-0.12$  & $2$&$.66$ &$\pm0.87   $&$+0.27$  & $-0.24$ \\[3pt]
$8$&$-$&$12$  & $1$&$.24$  &$\pm0.22   $&$+0.08$  & $-0.08$  & $2$&$.21$ &$\pm0.45   $&$+0.26$  & $-0.22$ \\[3pt]
$12$&$-$&$18$ & $0$&$.85$  &$\pm0.14   $&$+0.06$  & $-0.06$  & $1$&$.61$ &$\pm0.28   $&$+0.15$  & $-0.15$ \\[3pt]
$18$&$-$&$30$ & $0$&$.320$ &$\pm0.058  $&$+0.026$ & $-0.022$ & $0$&$.63$ &$\pm0.12   $&$+0.09$  & $-0.08$  \\[3pt]
$30$&$-$&$40$ & $0$&$.262$ &$\pm0.055  $&$+0.019$ & $-0.017$ & $0$&$.57$ &$\pm0.10   $&$+0.05$  & $-0.04$  \\[3pt]
$40$&$-$&$50$ & $0$&$.234$ &$\pm0.049  $&$+0.020$ & $-0.019$ & $0$&$.507$ &$\pm0.093 $&$+0.040$ & $-0.036$ \\[3pt]
$50$&$-$&$80$ & $0$&$.077$ &$\pm0.017  $&$+0.007$ & $-0.007$ & $0$&$.153$ &$\pm0.030$ &$+0.016$ & $-0.016$ \\[3pt]
$80$&$-$&$180$& $0$&$.0084$&$\pm0.0026$ &$+0.0006$& $-0.0005$ & $0$&$.0150$&$\pm0.0036$&$+0.0010$& $-0.0009$\\[3pt]
        \hline
        \hline
    \end{tabular}
    \label{tab:dsdqt}
\end{table}

\begin{table}[htpb]
  \centering
  \caption{Measured diphoton differential cross section  as a function of
    $\Delta\phi_{\Pgg\Pgg}$ for the two photon pseudorapidity ranges, with statistical (stat.) and systematic (syst.) uncertainties.}
  \vspace{0.5cm}
   \begin{tabular}{r@{}c@{}l|r@{}l @{\hspace{1em}} l@{\hspace{1em}}l@{\ \ }l | r@{}l @{\hspace{1em}} ll@{\ \ }l}
    \hline
    \hline\\[-.3cm]
    \multicolumn{13}{c}{\Large$d\sigma/d\Delta\phi_{\Pgg\Pgg}$ [pb]}\\[3pt]
    \hline
    \multicolumn{3}{c|}{$\Delta\phi_{\Pgg\Pgg}$}& \multicolumn{5}{c|}{$|\eta| < 1.44$} & \multicolumn{5}{c}{$|\eta| < 2.5$}\\[3pt] \hline
    & & &       &     &\multicolumn{1}{c}{stat.}&\multicolumn{2}{c|}{syst.}           &         &   &\multicolumn{1}{c}{stat.}   &\multicolumn{2}{c}{syst.}\\[3pt]
    $0$&$-$&$0.2\pi$         & $1$ &$.87$&$\pm0.53 $&$+0.13$& $-0.13$& $4$  &$.65$&$\pm0.89$&$+0.29$& $-0.30$ \\[3pt]
    $0.2\pi$&$-$&$0.4\pi$    & $1$ &$.77$&$\pm0.55 $&$+0.15$& $-0.14$& $5$  &$.5$ &$\pm1.1 $&$+0.5$ & $-0.4$  \\[3pt]
    $0.4\pi$&$-$&$0.6\pi$    & $3$ &$.09$&$\pm0.72 $&$+0.31$& $-0.29$& $5$  &$.5$ &$\pm1.3 $&$+0.6$ & $-0.5$  \\[3pt]
    $0.6\pi$&$-$&$0.8\pi$    & $7$ &$.2$ &$\pm1.1  $&$+0.5$ & $-0.4$ & $16$ &$.1$ &$\pm2.1 $&$+1.4$ & $-1.2$ \\[3pt]
    $0.8\pi$&$-$&$0.88\pi$   & $20$&$.8$ &$\pm2.6  $&$+1.0$ & $-1.0$ & $36$ &$.7$ &$\pm5.3 $&$+3.4$ & $-3.0$ \\[3pt]
    $0.88\pi$&$-$&$0.92\pi$  & $29$&$.8$ &$\pm5.1  $&$+1.7$ & $-1.5$ & $67$ &     &$\pm11  $&$+5$   & $-5$ \\[3pt]
    $0.92\pi$&$-$&$0.95\pi$  & $36$&$.2$ &$\pm8.1  $&$+5.1$ & $-4.7$ & $66$ &     &$\pm15  $&$+9$   & $-8$ \\[3pt]
    $0.95\pi$&$-$&$0.98\pi$  & $58$&$.8$ &$\pm8.8  $&$+4.2$ & $-3.8$ & $103$&     &$\pm17  $&$+12$  & $-11$ \\[3pt]
    $0.98\pi$&$-$&$\pi$      & $68$&     &$\pm11   $&$+4$   & $-4$   & $141$&     &$\pm23  $&$+12$  & $-11$ \\[3pt]
    \hline
    \hline
  \end{tabular}
  \label{tab:dsdphi}
\end{table}

\begin{table}[htpb]
  \centering
  \caption{Measured diphoton differential cross section as a function of
   $|{\cos\theta^*}|$ for the two photon pseudorapidity ranges, with statistical (stat.) and systematic (syst.) uncertainties.}
  \vspace{0.5cm}
   \begin{tabular}{r@{}c@{}l|r@{}l @{\hspace{1em}} l@{\hspace{1em}}l@{\ \ }l | r@{}l @{\hspace{1em}} ll@{\ \ }l}
    \hline
    \hline\\[-.3cm]
    \multicolumn{13}{c}{\Large$d\sigma/d|{\cos\theta^*}|$ [pb]}\\[3pt]
    \hline
    \multicolumn{3}{c|}{$|{\cos\theta^*}|$}& \multicolumn{5}{c|}{$|\eta| < 1.44$} & \multicolumn{5}{c}{$|\eta| < 2.5$}\\[3pt]
\hline
    & & &       &     &\multicolumn{1}{c}{stat.}&\multicolumn{2}{c|}{syst.}           &         &   &\multicolumn{1}{c}{stat.}   &\multicolumn{2}{c}{syst.}\\[3pt]
    $0$&$-$&$0.2$   & \hspace{1em}$52$&$.6$ &$\pm5.2  $&$+3.1$ & $-3.2$ & \hspace{1em}$87$&$.3$ &$\pm9.0  $&$+9.1$ & $-7.9$ \\[3pt]
    $0.2$&$-$&$0.4$ & $38$&$.4$ &$\pm4.9  $&$+3.0$ & $-3.0$ & $67$&$.0$ &$\pm8.2  $&$+6.6$ & $-6.0$ \\[3pt]
    $0.4$&$-$&$0.6$ & $34$&$.8$ &$\pm4.6  $&$+2.7$ & $-2.5$ & $66$&$.0$ &$\pm7.5  $&$+5.9$ & $-5.3$ \\[3pt]
    $0.6$&$-$&$0.8$ & $25$&$.6$ &$\pm3.7  $&$+1.6$ & $-1.5$ & $66$&$.7$ &$\pm7.7  $&$+6.1$ & $-5.3$ \\[3pt]
    $0.8$&$-$&$1$   & $6$ &$.4$ &$\pm1.4  $&$+0.3$ & $-0.4$ & $30$&$.8$ &$\pm7.9  $&$+5.9$ & $-4.7$ \\[3pt]
    \hline
    \hline
  \end{tabular}
  \label{tab:xsec_costhetas}
\end{table}

\section{Summary}

The integrated and differential production cross sections for isolated
photon pairs have been measured in proton-proton collisions at a
centre-of-mass energy of $7\TeV$, using data collected by the CMS
detector in 2010, corresponding to an integrated luminosity of
$36\pbinv$. The differential cross sections have been
measured as functions of the diphoton invariant mass, the diphoton
transverse momentum, the difference between the two photon azimuthal
angles, and $|{\cos\theta^*}|$.
The background from hadron decay products was estimated with a
statistical method based on an electromagnetic energy isolation
variable $\mathcal{I}$. The signal and background distributions for
$\mathcal{I}$ were entirely extracted from data, resulting in
systematic uncertainties of approximately $10\%$ on the measured
diphoton yields.

The measurements have been compared to a theoretical prediction
performed at next-to-leading-order accuracy using the state-of-the-art
fixed-order computations~\cite{Binoth:1999qq,Bern:2002jx}. Whereas
there is an overall agreement between theory and data for the diphoton
mass spectrum, the theory underestimates the cross section in
regions of the phase space where the two photons have an azimuthal
angle difference $\Delta\varphi\lesssim 2.8$.

\section*{Acknowledgements}

We wish to express our gratitude to J-Ph.~Guillet, E.~Pilon, Z.~Bern,
L.~Dixon, and C.~Schmidt for the fruitful discussions on the
theoretical aspects concerning the measurement.

We wish to congratulate our colleagues in the CERN accelerator
departments for the excellent performance of the LHC machine. We thank
the technical and administrative staff at CERN and other CMS
institutes, and acknowledge support from: FMSR (Austria); FNRS and FWO
(Belgium); CNPq, CAPES, FAPERJ, and FAPESP (Brazil); MES (Bulgaria);
CERN; CAS, MoST, and NSFC (China); COLCIENCIAS (Colombia); MSES
(Croatia); RPF (Cyprus); Academy of Sciences and NICPB (Estonia);
Academy of Finland, MEC, and HIP (Finland); CEA and CNRS/IN2P3
(France); BMBF, DFG, and HGF (Germany); GSRT (Greece); OTKA and NKTH
(Hungary); DAE and DST (India); IPM (Iran); SFI (Ireland); INFN
(Italy); NRF and WCU (Korea); LAS (Lithuania); CINVESTAV, CONACYT,
SEP, and UASLP-FAI (Mexico); MSI (New Zealand); PAEC (Pakistan); SCSR
(Poland); FCT (Portugal); JINR (Armenia, Belarus, Georgia, Ukraine,
Uzbekistan); MST, MAE and RFBR (Russia); MSTD (Serbia); MICINN and
CPAN (Spain); Swiss Funding Agencies (Switzerland); NSC (Taipei);
TUBITAK and TAEK (Turkey); STFC (United Kingdom); DOE and NSF (USA).
Individuals have received support from the Marie-Curie programme and
the European Research Council (European Union); the Leventis
Foundation; the A. P. Sloan Foundation; the Alexander von Humboldt
Foundation; the Belgian Federal Science Policy Office; the Fonds pour
la Formation \`a la Recherche dans l'Industrie et dans l'Agriculture
(FRIA-Belgium); the Agentschap voor Innovatie door Wetenschap en
Technologie (IWT-Belgium); and the Council of Science and Industrial
Research, India.

\newpage
\bibliography{auto_generated}   
\cleardoublepage \appendix\section{The CMS Collaboration \label{app:collab}}\begin{sloppypar}\hyphenpenalty=5000\widowpenalty=500\clubpenalty=5000\textbf{Yerevan Physics Institute,  Yerevan,  Armenia}\\*[0pt]
S.~Chatrchyan, V.~Khachatryan, A.M.~Sirunyan, A.~Tumasyan
\vskip\cmsinstskip
\textbf{Institut f\"{u}r Hochenergiephysik der OeAW,  Wien,  Austria}\\*[0pt]
W.~Adam, T.~Bergauer, M.~Dragicevic, J.~Er\"{o}, C.~Fabjan, M.~Friedl, R.~Fr\"{u}hwirth, V.M.~Ghete, J.~Hammer\cmsAuthorMark{1}, M.~Hoch, N.~H\"{o}rmann, J.~Hrubec, M.~Jeitler, W.~Kiesenhofer, A.~Knapitsch, M.~Krammer, D.~Liko, I.~Mikulec, M.~Pernicka$^{\textrm{\dag}}$, B.~Rahbaran, H.~Rohringer, R.~Sch\"{o}fbeck, J.~Strauss, A.~Taurok, F.~Teischinger, C.~Trauner, P.~Wagner, W.~Waltenberger, G.~Walzel, E.~Widl, C.-E.~Wulz
% ...
\vskip\cmsinstskip
\dag:~Deceased\\
1:~~Also at CERN, European Organization for Nuclear Research, Geneva, Switzerland\\
% ...

\end{sloppypar}
\end{document}